\documentclass[%
 aip,
 sd,%
 amsmath,amssymb,
 reprint,%
]{revtex4-1}

\usepackage{graphicx}% Include figure files
\usepackage{dcolumn}% Align table columns on decimal point
\usepackage{bm}% bold math

\begin{document}

%\preprint{}

\title{Resonant terms on gyro-center coordinate introduced by resonant electromagnetic perturbations} %Title of paper



\author{Shuangxi Zhang}
\email[]{zshuangxi@gmail.com; zhang.shuangxi.3s@kyoto-u.ac.jp}
%\thanks{}
%\affiliation{Institute of Fluid Physics, China Academy of Engineering Physics, Mianyang 621999, China.}
\affiliation{Graduate School of Energy Science, Kyoto University, Uji, Kyoto 611-0011, Japan.}

%\author{Shunagxi Zhang}
%\email[]{}
%%\homepage[]{Your web page}
%%\thanks{}
%%\altaffiliation{}
%\affiliation{}

%\author{J.Q.Li}
%%\email[]{Your e-mail address}
%%\homepage[]{Your web page}
%%\thanks{}
%%\altaffiliation{}
%\affiliation{Southwestern Institute of Physics, PO Box 432, Chengdu 610041, People¡¯s Republic of China}
%
%\author{Y.Kishimoto}
%%\email[]{Your e-mail address}
%%\homepage[]{Your web page}
%%\thanks{}
%%\altaffiliation{}
%\affiliation{Graduate School of Energy Science, Kyoto University, Uji, Kyoto 611-0011, Japan}


\date{\today}

\begin{abstract}
It's pointed out that the treatment of the resonant electromagnetic perturbation with the Lie transform method adopted in the gyrokinetic theory generates some nonphysical terms in the trajectory equations. By utilizing a modified application of this transform method, the resonant terms  in the trajectory equations satisfying each resonant condition are found out up to the second harmonic resonance. Through separating the fast-dynamic terms from the slow-dynamic ones, the slow-dynamic trajectory equations including the resonant terms are derived for various resonant conditions. The slow-dynamic evolution equation of gyroangle reveals that the real gyrating frequency of the charged particle around the magnetic field line under driving by the resonant wave has a shift from the cyclotron frequency without the wave driving. To study the effect caused by the frequency shift, a simple example for the fundamental-frequency cyclotron resonance is presented.
\end{abstract}

%\pacs{}% insert suggested PACS numbers in braces on next line

\maketitle %\maketitle must follow title, authors, abstract and \pacs

\section{INTRODUCTION}\label{sec1}

The radio frequency (RF) wave heating is a fundamental auxiliary method to heat the bulk plasma to the fusion temperature\cite{1975stix, wesson2004, 1987fisch, stixbook, ffchenbook, swansonbook, 1995philips, 1994erckmannppcf, 1981hwang, 1977nfperkins}. There are two basic processes of RF wave in plasma: one is the mode conversion of the incident wave to another propagating wave in plasma; the other one is the wave-particle resonant interaction. The former one is responsible for the energy exchange between the different waves. Except for collision, the dissipation of energy of wave must be through the wave-particle resonant interaction to transfer the energy of the wave to the charged particles. The wave-particle resonant interaction includes Landau resonance and cyclotron resonance\cite{wesson2004, ffchenbook, stixbook}. The heating scheme in tokamak plasma usually uses the fundamental cyclotron frequency resonance for the minority heating and the second harmonic resonance for the majority heating\cite{wesson2004, ffchenbook, stixbook}.

One simple way to calculate the motion of the charged particle under the action of the resonant electromagnetic wave is to calculate the full physical orbit advanced by the Lorentz force. But this method is extremely time consuming.
The other way is to separate the slow dynamics from the fast one by the gyro-center transform (GT).
The commonly applied GT adopts Cary-Littlejohn single-parameter Lie perturbation theory (CLSLP) to decoupling gyroangle $\theta$ from other degrees of freedom\cite{1983littlejohn,1990brizard,1988hahm,1988hahm2, 2007brizard1, 2009cary, 1999honqingpop, 2006shaojiewang, 2015tronko, 2016tronko}. The derived trajectory equations based on this method is widely used in the numerical simulation\cite{zhlinpop1995, Idomuranf2009, wwleejcp1987, xuxueq1991, biancalanipop2016}. Compared with non-cyclotron-resonant wave, the cyclotron resonant wave could introduce the secularity property to the gauge function $S$.
To prevent this secularity, Ref.(\onlinecite{gunyoungpop2007}) utilizes a method given in Ref.(\onlinecite{1983cary}) to remove the cyclotron resonant branch from the equation of gauge function. The original idea of the utility of this method in Ref.(\onlinecite{1983cary}) is to avoid the secularity of gauge function caused by the resonant perturbations, and is very similar to Von Zeipel method adopted in the canonical perturbation theory to remove the secularity of the generation function caused by the resonant perturbation \cite{abudullaev2006,lichtenberg1992}.
However, our inspection shows that the trajectory equations given by the Lagrangian 1-form derived from the foregoing treatment include some nonphysical terms, in particular, the nonphysical cyclotron resonant terms.

Instead, a modified application of CLSLP is adopted. The amplitude of each physical quantity for perturbation is formally given an order denoted by an independent parameter. Then, the new 1-form can be expanded order by order based on those independent parameters. For the application, those parameters can be replaced by the real values correspondingly. The truncation can be carried out at the designed order.

The resonant terms can be isolated out from those expanding terms based on the specific resonant conditions. In this paper, the resonant conditions are considered up to the second harmonic resonance. By getting rid of the fast-dynamic terms, a group of slow-dynamic trajectory equations are derived only including the effect of equilibrium field and the resonant contribution for each resonant condition. The  equation for gyroangle $\theta$ reveals that the gyrating frequency of charged particle under driving by the resonant wave is different from the original cyclotron frequency. This frequency shift would cause the optimal cyclotron resonant driving frequency to deviate from the original cyclotron frequency. This is proved by a simple numerical example presented in the context.

The arrangement of the remaining  paper is as follows.
Sec.\ref{sec11} repeats the calculating with CLSLP for the resonant perturbation.
Sec.\ref{sec3} gives some comments about the results given in Sec.(\ref{sec11}).
In Sec.\ref{sec10}, a modified application of CLSLP is adopted to expanding the new 1-form and isolating out those resonant terms for the different resonant conditions.
Sec.\ref{sec13} presents an example to show the shift of resonant frequency for the fundamental-frequency cyclotron resonance.
Sec.\ref{sec8} is  summary and discussion.

\section{Carrying out resonant CLSLP over guiding center Lagrangian 1-form including REW}\label{sec11}

The purpose GT pursues is to find a near identical transform (NIT) of coordinate with the gyro-angle $\theta$ in the new coordinate decoupled from other degrees of freedom up to some order of the small parameter which characterizes the amplitude of perturbation.

\subsection{Normalizing physical quantities}\label{sec2.1}

The basic physical quantity will be used is the Lagrangian differential 1-form of the motion of charged particle chosen from the magnetized plasma with the equilibrium magnetic potential
\begin{equation}\label{a119}
\gamma ' = \left( {q{\bf{A}}\left( {{{\bf{x}}}} \right) + m{\bf{v}}} \right)\cdot d{\bf{x}} - \frac{1}{2}m{v^2}dt.
\end{equation}
$(\mathbf{x},\mathbf{v})$ is the full physical coordinate frame.
By decoupling the gyroangle $\theta$ from other degrees of freedom up to $O(\varepsilon^2)$ with $\varepsilon  = \frac{\bm{\rho} }{{{L_0}}}$, it gives the guiding center Lagrangian 1-form like
\begin{eqnarray}\label{g88}
{\gamma _0} = && \left( {q{\bf{A}}\left( {{{\bf{X}}_{\bf{1}}}} \right) + m{U_1}{\bf{b}}} \right)\cdot d{{\bf{X}}_1} + \frac{m}{q}{\mu _1}d{\theta _1} \nonumber \\
&&- ({\mu _1}B\left( {{{{\bf{X}}}_{{1}}}} \right) + \frac{1}{2}m U_1^2)dt,
\end{eqnarray}
where $\mathbf{A}(\mathbf{X}_1)$ is the equilibrium magnetic potential. Then, differential 1-form for the perturbation wave is introduced
\begin{equation}\label{g89}
{\gamma _w} = q{{\bf{A}}_w}({{\bf{X}}_1} + {\bm{\rho }},t)\cdot d\left( {{{\bf{X}}_1} + {\bm{\rho }}} \right) - q\phi_w \left( {{{\bf{X}}_1} + {\bm{\rho }},t} \right)dt,
\end{equation}
with ${\bm{\rho }} = {{\bm{\rho }}'_0} + ( \cdots )$ and ${\bm{\rho}' _0} = \frac{1}{q}\sqrt {\frac{{2m{\mu _1}}}{{B\left( {{{\bf{X}}_1}} \right)}}} \left( { - {{\bf{e}}_1}\cos \theta  + {{\bf{e}}_2}\sin \theta } \right)$. $(\mathbf{b},\mathbf{e}_1,\mathbf{e}_2)$ forms a right-hand cartesian coordinate frame, $\mathbf{b}$ is the unit vector directing along equilibrium magnetic field $\mathbf{B}(\mathbf{X})$, and $\theta$ is the gyroangle. The rotation direction of ions around the magnetic field line is inverse to that of electrons.
Symbol $``(\cdots)"$ means higher order terms. $\mathbf{A}_w,\phi_w$ denotes the perturbations of the magnetic potential and the electric potential, respectively. Here, $\mathbf{A}(\mathbf{X_1})$ is the equilibrium magnetic potential, and  the guiding center coordinates plus the time is denoted as ${{\bf{Z}}}_1 = ({{\bf{X}}}_1,{U}_1,{\mu}_1 ,{\theta}_1, t)$. The other notations in Eqs.(\ref{g88},\ref{g89}) can be referred to Ref.(\onlinecite{1990brizard}).

The test particle is chosen from a thermal equilibrium plasma ensemble, e.g., the thermal equilibrium plasma in tokamak. Therefore, $\mathbf{A},U_1,\mathbf{X}_1,t,\mathbf{B}, \phi_w, \mu_1$ can be normalized by $A_0=B_0 L_0,v_t,L_0,L_0/v_t,B_0,A_0 v_t, mv_t^2/B_0$, respectively. $B_0, L_0$ are characteristic amplitude and spatial length of the magnetic field, respectively. $v_t$ is the thermal velocity of the particle ensemble which contains the test particle. The small parameter representing the normalized amplitude of $\mathbf{A}_w,\phi_w$ is extracted out, so that $\mathbf{A}_w,\phi_w$ are reformulated as $\varepsilon_w \mathbf{A}_w, \varepsilon_w \phi_w$, respectively, with $O(|\mathbf{A}_w|)\sim O(|\phi_w|)\sim O(1)$. Throughout the rest of the paper, all physical quantities are normalized.

The detailed normalization procedure is given by taking Eq.(\ref{g88}) as an example. First, divide both sides of Eq.(\ref{g88}) by $m{v_t}{L_0}$. The first term of RHS of Eq.(\ref{g88}) is like $\frac{{q{A_0}}}{{m{v_t}}}\frac{{{\bf{A}}\left( {{{\bf{X}}_1}} \right)}}{{{A_0}}}\cdot\frac{{d{{\bf{X}}_1}}}{{{L_0}}}$, which is further written as $\frac{1}{\varepsilon }{\bf{A}}\left( {{{\bf{X}}_1}} \right)\cdot d{{\bf{X}}_1}$, with the replacement: $\varepsilon  \equiv \frac{{m{v_t}}}{{q{A_0}}}$, $\frac{{{\bf{A}}\left( {{{\bf{X}}_1}} \right)}}{{{A_0}}} \to {\bf{A}}\left( {{{\bf{X}}_1}} \right),\frac{{d{{\bf{X}}_1}}}{{{L_0}}} \to d{{\bf{X}}_1}$. Other terms can be normalized in the same way. Eventually, we could derive a normalized Lagrangian 1-form like
\begin{eqnarray}
\frac{{{\gamma _0}}}{{m{v_t}{L_0}}} =& \left( {\frac{1}{\varepsilon}{\bf{A}}\left( {{{\bf{X}}_1}} \right) + {U_1}{\bf{b}}} \right)\cdot d{{\bf{X}}_1} + \varepsilon{\mu _1}d{\theta _1} \nonumber \\
& - ({\mu _1}B\left( {{{\bf{X}}_1}} \right) + \frac{1}{2}U_1^2)dt,
\end{eqnarray}
by utilizing the normalization scheme given previously. Now, multiplying both sides by $\varepsilon$, and rewriting $\frac{{{\varepsilon \gamma _0}}}{{m{v_t}{L_0}}}$ to be $\gamma_0$, we derive the normalized 1-form as follows
\begin{eqnarray}\label{g2}
{\gamma _0} =&& {\bf{A}}\left( {{{\bf{X}}_1}} \right)\cdot d{{\bf{X}}_1} + \varepsilon {U_1}{\bf{b}}\cdot d{{\bf{X}}_1} + {\varepsilon ^2}{\mu _1}d{\theta _1} \nonumber \\
&& - \varepsilon \left( {\frac{{U_1^2}}{2} + {\mu _1}B\left( {{{\bf{X}}_1}} \right)} \right)dt.
\end{eqnarray}
Since a constant factor $\frac{\varepsilon }{{m{v_t}{L_0}}}$ doesn't change the dynamics determined by the Lagrangian 1-form, the Lagrangian 1-form given by Eq.(\ref{g2}) is of the same dynamics with that given by Eq.(\ref{g88}).
By utilizing the same normalization procedure, with $\mathbf{A}_w,\phi_w$ changed to be $\varepsilon_w \mathbf{A}_w, \varepsilon_w\phi_w$, respectively, Eq.(\ref{g89}) becomes
\begin{eqnarray}\label{g3}
&&{\gamma _w} = \varepsilon_w {{\bf{A}}_w}\left( {{{\bf{X}}_1} + \bm{\rho} ,t} \right)\cdot d\left( {{{\bf{X}}_{\bf{1}}} + \bm{\rho} } \right) + \varepsilon_w {\phi _w}\left( {{{\bf{X}}_{\bf{1}}} + \bm{\rho} ,t} \right)dt \nonumber \\
&& \approx \varepsilon_w \exp \left( {{\varepsilon \bm{\rho} _0}\cdot\nabla } \right){{\bf{A}}_w}\left( {{{\bf{X}}_1},t} \right)\cdot\left( \begin{array}{l}
d{{\bf{X}}_1} + \frac{{\varepsilon\partial { \bm{\rho} _0}}}{{\partial {{\bf{X}}_1}}}\cdot d{{\bf{X}}_1} \nonumber \\
 + \frac{{\varepsilon \partial {{\bm{\rho}} _0}}}{{\partial {\mu _1}}}d{\mu _1} + \frac{{\varepsilon \partial { \bm{\rho} _0}}}{{\partial {\theta _1}}}d{\theta _1}
\end{array} \right)  \nonumber \\
&&  - \varepsilon_w \exp \left( {{\varepsilon \bm{\rho} _0}\cdot\nabla } \right){\phi _w}\left( {{{\bf{X}}_1},t} \right)dt,
\end{eqnarray}
where ${\varepsilon } \equiv \frac{{m{v_t}}}{{{A_0}q}}=\frac{{{\bm{\rho} _t}}}{{{L_0}}}$, ${\bm{\rho} _t} = \frac{{m{v_t}}}{{{B_0}q}}$, ${\bm{\rho} _0} = \sqrt {\frac{{2\mu }}{{B\left( {\bf{X}}_1 \right)}}} \left( { - {{\bf{e}}_1}\cos \theta  + {{\bf{e}}_2}\sin \theta } \right)$. In Eqs.(\ref{g2},\ref{g3}), all $\varepsilon,\varepsilon_w$ takes part in calculation. If the small parameters $\varepsilon,\varepsilon_w$ are just used as a symbol of the order of terms, they are denoted as $\varepsilon^*, \varepsilon_w^*$. This rule is adopted throughout the rest of this paper.

\subsection{Carrying out the pullback transform and deriving the trajectory equations}\label{sec2.2}

$\gamma_{0}+\gamma_{w}$ is the total Lagrangian differential 1-form with the fast angle $\theta$ included in $\bm{\rho}_0$.  To decouple $\theta$ from other degrees of freedom, CLSLP is adopted with $\varepsilon_w$ treated as the small order parameter, while $\varepsilon$ as a normal quantity not involved in the order expanding. The gyrocenter frame plus the time is recorded as ${\bf{Z}} = \left( {{\bf{X}},\mu ,U,\theta },t \right)$.
The coordinate transform should satisfy NIT and is formally recorded  as ${{\bf{Z}}_1} = \exp \left( { - {\varepsilon _w}{g^i_1}\left( {\bf{Z}} \right){\partial _i}} \right){\bf{Z}}$ with  $O(g^i_1)\sim O(1)$ for all $i\in \{\mathbf{X},U,\mu,\theta\}$. The time keeps the same before and after transform. All $g^i_1$s need to be solved. The new $\Gamma$ induced by this coordinate transform is
\begin{equation}\label{a99}
\Gamma=[\cdots T_2 T_1(\gamma_0+\gamma_w)](\mathbf{Z})+dS,
\end{equation}
with ${T_i} = \exp ( - {\varepsilon _w}{L_{{g^j_1}}}({\bf{Z}}){\partial _{{Z^j}}})$.
Expand $\Gamma$ in Eq.(\ref{a99}) as the following power series
\begin{equation}
\Gamma  = \sum\limits_{n\ge 0} {\frac{1}{{n!}}\varepsilon _w^n{\Gamma _i}}.
\end{equation}
$O(\varepsilon_w^0)$ part of new $\Gamma$ is
\begin{equation}\label{g35}
{\Gamma _0} = {{\bf{A}}}\left( {\bf{X}} \right)\cdot d{\bf{X}} + {\varepsilon }U{\bf{b}}\cdot d{\bf{X}} + {\varepsilon ^{2}}\mu d\theta
 - {\varepsilon }H_0 dt,
\end{equation}
with $H_0={\frac{{U_{}^2}}{2} + \mu B\left( {\bf{X}} \right)}$. The $O(\varepsilon_w)$ part is
%\begin{widetext}

\begin{eqnarray}\label{g36}
{\varepsilon _w}{\Gamma _1} =&& \left( { - \left( {{\bf{B}} + \varepsilon U\nabla  \times {\bf{b}}} \right) \times \left( {{\varepsilon _w}{{\bf{g}}_1^X}} \right) - \varepsilon {\varepsilon _w}{g_1^U}{\bf{b}} + \exp \left( {{\varepsilon \bm{\rho} _0}\cdot{\nabla}} \right)\left( {{\varepsilon _w}{{\bf{A}}_w}} \right)} \right)\cdot d{\bf{X}} \nonumber \\
&& + \varepsilon \left( {{\varepsilon _w}{{\bf{g}}_1^X}\cdot{\bf{b}}} \right)dU + \left( {\exp \left( {{\varepsilon \bm{\rho} _0}\cdot{\nabla}} \right)\left( {{\varepsilon _w}{{\bf{A}}_w}} \right)\cdot\frac{{\varepsilon \partial { \bm{\rho} _0}}}{{\partial \theta }} - {\varepsilon ^2 \varepsilon_w}{g_1^\mu }} \right)d\theta  \nonumber \\
&& - \left( {{\varepsilon ^2}\left( {{\varepsilon _w}{g_1^\theta }} \right) + \exp \left( {\varepsilon {\bm{\rho} _0}\cdot{\nabla}} \right)\left( {{\varepsilon _w}{{\bf{A}}_w}} \right)\cdot\frac{{\varepsilon\partial { \bm{\rho} _0}}}{{\partial \mu }}} \right)d\mu  \nonumber \\
&& - \left( { - \left( {{\varepsilon _w}{{\bf{g}}_1^X}\cdot\nabla {H_0}} \right) - \varepsilon {\varepsilon _w}U{g_1^U} - \varepsilon {\varepsilon _w}{g_1^\mu }B + \exp \left( {{\varepsilon \bm{\rho} _0}\cdot{\nabla}} \right)\left( {{\varepsilon _w}{\phi _w}} \right)} \right)dt
 + {\varepsilon _w}dS.
\end{eqnarray}

%\end{widetext}
Eq.(\ref{g36}) obviously shows the order confusion between $\varepsilon$ and $\varepsilon_w$.

Current GT assumes the following identities\cite{1990brizard}
\begin{equation}\label{g34}
\Gamma_{1i}=0,i\in (\mathbf{X},U,\mu,\theta).
\end{equation}
Then, all of $g^i$s can be derived based on Eqs.(\ref{g36},\ref{g34}) as
\begin{eqnarray}\label{g37}
{{\bf{g}}_1^X} = && - \frac{1}{{{\bf{b}}\cdot{{\bf{B}}^*}}}\left( {{\bf{b}} \times \exp \left( {\varepsilon \bm{\rho}_0 \cdot\nabla } \right){{\bf{A}}_w}\left( {{\bf{X}},t} \right)  + {\bf{b}} \times \nabla {S_1}} \right)  \nonumber \\
&&- \frac{{{{\bf{B}}^*}}}{{{\varepsilon }}}\frac{{\partial {S_1}}}{{\partial U}},
\end{eqnarray}
where ${{\bf{B}}^*} = {\bf{B}} + {\varepsilon }U\nabla  \times {\bf{b}}$.
\begin{equation}\label{g38}
{g_1^U} = \frac{1}{{{\varepsilon }}}{\bf{b}}\cdot\exp \left( {\varepsilon \bm{\rho}_0 \cdot\nabla } \right){{\bf{A}}_w}\left( {{\bf{X}},t} \right) + \frac{1}{{{\varepsilon }}}{\bf{b}}\cdot\nabla {S_1},
\end{equation}
\begin{equation}\label{g39}
{g_1^\mu } = \frac{1}{{{\varepsilon}}}\exp \left( {\varepsilon \bm{\rho}_0 \cdot\nabla } \right){{\bf{A}}_w}\left( {{\bf{X}},t} \right)\cdot\frac{{ \partial \bm{\rho}_0 }}{{\partial \theta }} + \frac{1}{{{\varepsilon ^{2}}}}\frac{{\partial {S_1}}}{{\partial \theta }},
\end{equation}
\begin{equation}\label{g40}
{g_1^\theta } =  - \frac{1}{{{\varepsilon}}}\exp \left( {\varepsilon \bm{\rho}_0 \cdot\nabla } \right){{\bf{A}}_w}\left( {{\bf{X}},t} \right)\cdot\frac{{\partial \bm{\rho}_0 }}{{\partial \mu }} - \frac{1}{{{\varepsilon ^{2}}}}\frac{{\partial {S_1}}}{{\partial \mu }}.
\end{equation}
\begin{eqnarray}\label{g41}
\frac{{\partial {S_1}}}{{\partial t}} + U{\bf{b}}\cdot\nabla {S_1} + \frac{B(\mathbf{X})}{{{\varepsilon }}}\frac{{\partial {S_1}}}{{\partial \theta }} = F + {\Gamma _{1t}},
\end{eqnarray}
with
\begin{eqnarray}\label{g92}
F = &&  \exp \left( {\varepsilon \bm{\rho}_0 \cdot\nabla } \right){\phi _w}\left( {{\bf{X}},t} \right) \nonumber \\
&& -  U{\bf{b}}\cdot\exp \left( {\varepsilon \bm{\rho}_0 \cdot\nabla } \right){{\bf{A}}_w}\left( {{\bf{X}},t} \right) \nonumber \\
&&-  {B(\mathbf{X})}\exp \left( {\varepsilon \bm{\rho}_0 \cdot\nabla } \right){{\bf{A}}_w}\left( {{\bf{X}},t} \right)\cdot\frac{{\partial \bm{\rho}_0 }}{{\partial \theta }}.
\end{eqnarray}
Smaller term ${\varepsilon_w {{\bf{g}}_1^X}\cdot \varepsilon \nabla {H_0}}$ and other higher order terms are ignored in Eq.(\ref{g41}).

For the low frequency perturbation, inequalities $\left| {\frac{{\partial {S_1}}}{{\partial t}}} \right| \ll \left| {\frac{B}{{{\varepsilon }}}\frac{{\partial {S_1}}}{{\partial \theta }}} \right|, \left| {U{\bf{b}}\cdot\nabla {S_1}} \right| \ll \left| {\frac{B}{{{\varepsilon }}}\frac{{\partial {S_1}}}{{\partial \theta }}} \right|$ hold, and the lowest order equation of Eq.(\ref{g41}) is
\begin{eqnarray}\label{g42}
\frac{B}{{{\varepsilon }}}\frac{{\partial {S_1}}}{{\partial \theta }} = F +\Gamma_{1t}.
\end{eqnarray}
To avoid the secularity of $S_1$ over the integration of $\theta$, $\Gamma_{1t}$ is chosen to be
\begin{equation}\label{g43}
{\Gamma_{1t}} = -\left\langle F \right\rangle,
\end{equation}
where $\left \langle F \right \rangle $ means the averaging over $\theta$.
The new $\Gamma$ approximated up to $O(\varepsilon_w)$ is
\begin{equation}\label{a100}
\Gamma  = \left( {{\bf{A}}\left( {\bf{X}} \right) + \varepsilon U{\bf{b}}} \right)\cdot d{\bf{X}} + {\varepsilon ^2}\mu d\theta  - \left( {{H_0} - {\varepsilon _w}{\Gamma _{1t}}} \right)dt,
\end{equation}
with ${H_0} = \varepsilon \left( {\frac{{{U^2}}}{2} + \mu B\left( {\bf{X}} \right)} \right)$ and ${{\Gamma _{1t}}}$ given in Eq.(\ref{g43}).

For the high frequency perturbation satisfying the fundamental-frequency resonance $\omega\approx \Omega(\mathbf{X})$ with $\Omega(\mathbf{X})=B(\mathbf{X})/\varepsilon$ around the resonant region,  the inequalities
\begin{equation}\label{g44}
\left| {U{\bf{b}}\cdot\nabla {S_1}} \right| \ll \left| {\frac{{\partial {S_1}}}{{\partial t}}} \right|\sim \left| {\frac{B}{{{\varepsilon }}}\frac{{\partial {S_1}}}{{\partial \theta }}} \right|
\end{equation}
hold,  and the lowest order equation of Eq.(\ref{g41}) is
\begin{equation}\label{g45}
\frac{{\partial {S_1}}}{{\partial t}} + \frac{B}{{{\varepsilon }}}\frac{{\partial {S_1}}}{{\partial \theta }} = F + \Gamma_{1t}.
\end{equation}

In the resonant region, term $ -{B(\mathbf{X})} \exp \left( {\varepsilon \bm{\rho}_0 \cdot\nabla } \right)$ ${{\bf{A}}_w}\left( {{\bf{X}},t} \right) \cdot\frac{{\partial \bm{\rho}_0 }}{{\partial \theta }}$ included in $F$ can be divided into two parts as
\begin{eqnarray}
&& - {B(\mathbf{X})}{{\bf{A}}_{w}}\left( {{\bf{X}},t} \right)\cdot\frac{{ \partial \bm{\rho}_0 }}{{\partial \theta }} \nonumber  \\
&& - {B(\mathbf{X})}\exp \left( {\varepsilon \bm{\rho}_0 \cdot\nabla } \right)'{{\bf{A}}_w}\left( {{\bf{X}},t} \right)\cdot\frac{{\partial \bm{\rho}_0 }}{{\partial \theta }} \nonumber
\end{eqnarray}
where $'$ represents that $n=0$ term in the exponential expansion is deleted. The term $- {B(\mathbf{X})}{{\bf{A}}_w}\left( {{\bf{X}},t} \right)\cdot\frac{{\partial \bm{\rho}_0 }}{{\partial \theta }}$ is responsible for the fundamental frequency cyclotron resonance, but not the true cyclotron resonant term as pointed out in Sec.(\ref{sec3}). In the resonant region, the composite frequency of this term is almost zero. The time integral of the resonant branch is a secularity term.   To avoid the secularity of $S_1$ over the time integration, $\Gamma_{1t}$ is chosen to be
\begin{eqnarray}\label{g46}
 {\Gamma _{1t}} =&&  {B(\mathbf{X})}{{\bf{A}}_w}\left( {{\bf{X}},t} \right)\cdot\frac{{\partial \bm{\rho}_0 }}{{\partial \theta }}  \nonumber \\
&& - \left\langle {F + {B(\mathbf{X})}{{\bf{A}}_w}\left( {{\bf{X}},t} \right)\cdot\frac{{\partial \bm{\rho}_0 }}{{\partial \theta }}} \right\rangle.
\end{eqnarray}

Eventually, up to order $O(\varepsilon_w)$, the new $\Gamma$ becomes
\begin{equation}\label{g47}
\Gamma  = {{\bf{A}}}\left( {\bf{X}} \right)\cdot d{\bf{X}} + {\varepsilon }U{\bf{b}}\cdot d{\bf{X}} + {\varepsilon ^{2}}\mu d\theta  - Hdt.
\end{equation}
\begin{eqnarray}\label{g51}
H =&& {\varepsilon}{H_0} - {\Gamma _{1t}} \nonumber \\
 =&& \frac{{{\varepsilon }U_{}^2}}{2} + {\varepsilon }\mu B({\bf{X}}) - \varepsilon_w B({\bf{X}}){{\bf{A}}_w}\left( {{\bf{X}},t} \right)\cdot\frac{{\partial {\bm{\rho} _0}}}{{\partial \theta }} \nonumber \\
&& + \varepsilon_w\left\langle {F +  B({\bf{X}}){{\bf{A}}_w}\left( {{\bf{X}},t} \right)\cdot\frac{{\partial {\bm{\rho} _0}}}{{\partial \theta }}} \right\rangle.
\end{eqnarray}
The Lagrangian given by Eq.(\ref{g47}) is
\begin{equation}\label{g48}
L = {{\bf{A}}}\left( {\bf{X}} \right)\cdot\mathop {\bf{X}}\limits^.  + {\varepsilon}U{\bf{b}}\cdot\mathop {\bf{X}}\limits^.  + {\varepsilon ^{2}}\mu \dot \theta -H.
\end{equation}
The Euler-Lagrangian equation of each variable based on the Lagrangian given by Eq.(\ref{g48}) gives the equations of motion.

First, the $\bf{X}$ component of Euler-Lagrangian equation is $\frac{d}{{dt}}\frac{{\partial L}}{{\partial \mathop {\bf{X}}\limits^. }} = \frac{{\partial L}}{{\partial {\bf{X}}}}$. The following several formulas are needed
\begin{eqnarray}\label{g49}
\frac{{\partial L}}{{\partial \mathop {\bf{X}}\limits^. }} =&& {{\bf{A}}}({\bf{X}}) + 2{\varepsilon }U{\bf{b}} - {\varepsilon }{\bf{b}}\frac{{\partial H}}{{\partial U}}  \nonumber \\
=&& {{\bf{A}}}({\bf{X}}) + {\varepsilon }U{\bf{b}},
\end{eqnarray}
\begin{equation}\label{g52}
\frac{d}{{dt}}\frac{{\partial L}}{{\partial \mathop {\bf{X}}\limits^. }} = \mathop {\bf{X}}\limits^. \cdot\nabla \left( {{{\bf{A}}(\mathbf{X})} + {\varepsilon }U{\bf{b}}} \right) + {\varepsilon }{\bf{b}}\frac{{dU}}{{dt}},
\end{equation}
\begin{equation}\label{g50}
\frac{{\partial L}}{{\partial {\bf{X}}}} = \nabla \left( {{{\bf{A}}(\mathbf{X})} + {\varepsilon }U{\bf{b}}} \right)\cdot\mathop {\bf{X}}\limits^.  - \nabla H.
\end{equation}
The $\bf{X}$ component of Euler-Lagrangian equation becomes
\begin{equation}\label{g53}
- \mathop {\bf{X}}\limits^.  \times {{\bf{B}}^*} + {\varepsilon }{\bf{b}}\frac{{dU}}{{dt}} =  - \nabla H.
\end{equation}
with ${{\bf{B}}^*} = \nabla  \times \left( {{{\bf{A}}}({\bf{X}}) + {\varepsilon }U{\bf{b}}} \right)$. By left cross multiplying Eq.(\ref{g53}) with $\bf{b}$, it's found that
\begin{equation}\label{g54}
\mathop {\bf{X}}\limits^. {\rm{ = }}\frac{{U{\bf{B}}^*(\mathbf{X})  + {\bf{b}} \times \nabla H}}{{{\bf{b}}\cdot{\bf{B}}^*}}.
\end{equation}
The dot product between $\mathbf{B}^*$ and Eq.(\ref{g53}) leads to
\begin{equation}\label{g55}
\dot{U} = \frac{{ - \left( {{\bf{B}}^*\cdot\nabla H} \right)}}{{{\varepsilon }{\bf{b}}\cdot{\bf{B}}^*}}.
\end{equation}

Second, the $\theta$ component of Euler-Lagrangian equation is $\frac{d}{{dt}}\frac{{\partial L}}{{\partial \dot \theta }} = \frac{{\partial L}}{{\partial \theta }}$. With $\frac{{\partial L}}{{\partial \dot \theta }} = {\varepsilon ^{2}}\mu $ and $\frac{{\partial L}}{{\partial \theta }} = \varepsilon_w {B(\mathbf{X})}{{\bf{A}}_w}\left( {{\bf{X}},t} \right)\cdot\frac{{{\partial ^2}\bm{\rho}_0 }}{{\partial {\theta ^2}}}$, it's derived that
\begin{equation}\label{g56}
\dot{\mu} = \frac{{{\varepsilon_w B(\mathbf{X})}}}{{{\varepsilon ^{2}}}}{{\bf{A}}_{w\perp}}\left( {{\bf{X}},t} \right)\cdot\frac{{{\partial ^2}\bm{\rho}_0 }}{{\partial {\theta ^2}}}.
\end{equation}

Third, the $\mu$ component of Euler-Lagrangian equation is $\frac{d}{{dt}}\frac{{\partial L}}{{\partial \dot \mu }} = \frac{{\partial L}}{{\partial \mu }}$. With $\frac{{\partial L}}{{\partial \dot \mu }} = 0$ and
\begin{eqnarray}\label{g57}
\frac{{\partial L}}{{\partial \mu }} =&& {\varepsilon ^{2}}\dot \theta  - {\varepsilon }B\left( {\bf{X}} \right) +\varepsilon_w B\left( {\bf{X}} \right){{\bf{A}}_{w\perp}}\left( {{\bf{X}},t} \right)\cdot\frac{{{\partial ^2}\bm{\rho}_0 }}{{\partial \mu \partial \theta }}  \nonumber \\
 && - \varepsilon_w \frac{\partial }{{\partial \mu }}\left\langle {F + {B(\mathbf{X})}{{\bf{A}}_{w\perp}}\left( {{\bf{X}},t} \right)\cdot\frac{{\partial \bm{\rho}_0 }}{{\partial \theta }}} \right\rangle,
\end{eqnarray}
it's derived that
\begin{eqnarray}\label{g58}
\dot \theta  = \frac{{B({\bf{X}})}}{\varepsilon } - \frac{{{\varepsilon _w}B({\bf{X}})}}{{{\varepsilon ^2}}}{{\bf{A}}_{w\perp}}\left( {{\bf{X}},t} \right)\cdot\frac{{{\partial ^2}{\bm{\rho} _0}}}{{\partial \mu \partial \theta }}  \nonumber \\
 + \frac{{{\varepsilon _w}}}{{{\varepsilon ^2}}}\frac{\partial }{{\partial \mu }}\left\langle {F + B({\bf{X}}){{\bf{A}}_{w\perp}}\left( {{\bf{X}},t} \right)\cdot\frac{{\partial {\bm{\rho} _0}}}{{\partial \theta }}} \right\rangle .
\end{eqnarray}

\section{Trajectory equations not consistent with real physics}\label{sec3}

In Eq.(\ref{g54})and (\ref{g55}), the contributions of the perturbations are mainly from the terms $\frac{{U{\bf{b}} \times \nabla {{\bf{A}}_{{w}}}}}{{{\bf{b}}\cdot{{\bf{B}}^*}}}$
and $\frac{{U{{\bf{B}}^*}}}{{{\varepsilon}B_\parallel ^*}}\cdot\nabla {{\bf{A}}_{{w}}}$, respectively. These terms are nonphysical, since $\mathbf{A}_w$ includes an arbitrary gauge term like $\nabla f$ with $f$ being an arbitrary function, and the term proportional to $\nabla \nabla f$ can't be cancelled.  The real physical contribution should be like $\frac{{\partial {{\bf{A}}_{\bf{w}}}/\partial t \times {\bf{b}}}}{{{\bf{b}}\cdot{B^*}}}$ and $- {\bf{b}}\cdot\frac{\partial }{{\partial t}}{{\bf{A}}_w}$, which are the $\mathbf{E}\times \mathbf{B}$ drift produced by the inductive electric field, and the parallel inductive electric field acceleration of charged particle, respectively.

The error of motion equation of $\mu$ given in Eq.(\ref{g56}) is not very easy to be observed. To make it exposure, we first construct a rectangular coordinate frame. As the usual way, the unit vector of the gyro radius may be written as \cite{1990brizard,1999honqingpop}
\begin{equation}\label{g59}
\hat {\bm{\rho} }_0 = -{{\mathbf{e}}_1}\cos \theta  + {{\mathbf{e}}_2}\sin \theta
\end{equation}
where unit vectors $\mathbf{e}_1$ and $\mathbf{e}_2$ with $\mathbf{e}_1\perp \mathbf{e}_2$ are locally perpendicular to the unit parallel magnetic vector $\mathbf{b(\mathbf{x})}$. Symbol $\widehat {( \cdots )}$ represents unit vector, and $\theta$ is the gyro angle. The non-normalized Larmor radius vector is $\bm{\rho}  = \frac{{m{v_ \bot }}}{{qB\left( {\bf{X}} \right)}}{\hat {\bm{\rho}} _0}$. As given before, the normalized edition of $\bm{\rho}$ is
%\begin{equation}\label{g64}
${\varepsilon }\bm{\rho}_0$ with ${\bm{\rho} _0} = \sqrt {\frac{{2\mu }}{{B\left( {\bf{X}} \right)}}} \left( { - {{\bf{e}}_1}\cos \theta  + {{\bf{e}}_2}\sin \theta } \right)$.
%\end{equation}
The unit perpendicular velocity vector is
\begin{equation}\label{g65}
{{{\mathbf{\hat v}}}_ \bot } = {{\mathbf{e}}_1}\sin \theta+{{\mathbf{e}}_2}\cos \theta.
\end{equation}
Then, the following relations can be derived
\begin{equation}\label{g66}
\frac{{\partial {{\hat {\bm{\rho} }}_0}}}{{\partial \theta }} = {\widehat {\bf{v}}_ \bot },\frac{{\partial {{\widehat {\bf{v}}}_ \bot }}}{{\partial \theta }} =  - {\hat {\bm{\rho}} _0}.
\end{equation}
Then, Eq.(\ref{g56}) can be rewritten as
\begin{equation}\label{g67}
\dot \mu  =- \frac{\varepsilon_w B(\bf{X})}{{{\varepsilon ^{2}}}}\sqrt {\frac{2\mu }{{B(\mathbf{X})}}} {{\bf{A}}_{w\perp}}\cdot\hat {\bm{\rho}}_0.
\end{equation}
% where Eq.(\ref{g64}) is applied and higher order terms are ignored.

Now we try to derive an approximate evolution equation of the magnetic moment simply from ${m}\frac{{d{}}}{{dt}} \mathbf{v}= q{\mathbf{v}} \times {\mathbf{B}} + q{{\mathbf{E}}_w}$ which can be normalized to be $\frac{{d{\bf{v}}}}{{dt}} = \frac{1}{\varepsilon }\left( {{\bf{v}} \times {\bf{B}} + {{\bf{E}}_w}} \right)$. The electric field is assumed to be circularly polarized.
The perpendicular velocity part responsible for the force imposed by the circularly polarized electric field can be extracted out as $\frac{{d{{\bf{v}}_ \bot }}}{{dt}} = \frac{1}{\varepsilon}{{\bf{E}}_w}$. Dot product both sides by $\frac{{{{\bf{v}}_ \bot }}}{{B\left( {\bf{x}} \right)}}$ to get $\frac{{{{\bf{v}}_ \bot }}}{{B\left( {\bf{x}} \right)}}\cdot\frac{{d{{\bf{v}}_ \bot }}}{{dt}} = \frac{{{{\bf{v}}_ \bot }}}{{\varepsilon B\left( {\bf{x}} \right)}}\cdot{{\bf{E}}_w}$. Noticing that ${v_ \bot } = \sqrt {2{\mu _1}B\left( {\bf{x}} \right)}$, we could approximately derive
\begin{eqnarray}\label{g70}
\frac{{d{\mu _1}}}{{dt}} &&=\frac{1}{{B\left( {\bf{x}} \right)\varepsilon }}{{\bf{v}}_ \bot }\cdot{{\bf{E}}_w} \nonumber \\
&& = \frac{1}{\varepsilon }\sqrt {\frac{{2\mu_1 }}{{B\left( {\bf{x}} \right)}}} {\widehat {\bf{v}}_ \bot }\cdot{{\bf{E}}_w}.
\end{eqnarray}

%Since $\omega\approx \varepsilon^{*{-1}}$, $O(|\mathbf{E}_w|)=O(\varepsilon^{*(\alpha-1)})$ stand, and all quantities of RHS of Eqs.(\ref{g67},\ref{g70}) are normalized, the order of RHS of Eq.(\ref{g70}) and  Eq.(\ref{g67}) are $O(\varepsilon^{*(\alpha-2)})$ and $O(\varepsilon^{*(\alpha-1)})$, respectively. Eq.(\ref{g70}) is obviously different from Eq.(\ref{g67}).

We give a specific example to explain the difference between Eq.(\ref{g67}) and (\ref{g70}). A test circular polarized electric field is introduced as
\begin{eqnarray}\label{g71}
{{\bf{E}}_w}({\bf{x}},t) = {E_w}\left( \begin{array}{l}
{{\bf{e}}_{1w}}\sin \left( {{\theta _w} - {\bf{k}}\cdot{\bf{x}}} \right)\\
 + {{\bf{e}}_{2w}}\cos \left( {{\theta _w} - {\bf{k}}\cdot{\bf{x}}} \right)
\end{array} \right)
\end{eqnarray}
where  the polarized angle is $\theta_w=\omega t+\theta_0$ with $\theta_0$  an arbitrary initial phase, and $E_w$ is the amplitude of the electric field. Unit vectors ${{{\mathbf{e}}_{1w}}}$ and $\mathbf{e}_{2w}$ with ${{{\mathbf{e}}_{1w}}}\perp \mathbf{e}_{2w}$ are perpendicular to the wave vector $\mathbf{k}$.

The normalized magnetic vector potential ${{{\bf{A}}_w}\left( {{\bf{X}},t} \right)}$ is derived from the Faraday's law ${{\bf{E}}_w} =  - \frac{\partial }{{\partial t}}{{\bf{A}}_{w}}$
\begin{eqnarray}\label{g72}
{{\bf{A}}_w}({\bf{x}},t) = \frac{{{E_w}}}{\omega }\left( \begin{array}{l}
 - {{\bf{e}}_{1w}}\cos ({\theta _w} - {\bf{k}}\cdot{\bf{x}})\nonumber \\
 + {{\bf{e}}_{2w}}\sin ({\theta _w} - {\bf{k}}\cdot{\bf{x}})
\end{array} \right).
\end{eqnarray}
Supposing $\mathbf{b}\parallel \mathbf{k}$ and $\mathbf{e}_1=\mathbf{e}_{1w},\mathbf{e}_2=\mathbf{e}_{2w}$, Eq.(\ref{g67}) and Eq.(\ref{g70}) become
\begin{equation}\label{g73}
\dot \mu  =- \frac{\varepsilon_w B(\mathbf{X})}{{{\varepsilon^2}}}\sqrt {\frac{2\mu }{{B(\mathbf{X})}}} \frac{{{E_w}}}{w}\cos \left( {{\theta _w} - \theta  - {\bf{k}}\cdot{\bf{X}}} \right),
\end{equation}
\begin{equation}\label{g74}
{{\dot \mu }_1} =\frac{{\varepsilon_w {E_w}}}{{{\varepsilon}}}\sqrt {\frac{{{2\mu _1}}}{{B\left( {{{\bf{X}}_1}} \right)}}} \cos \left( {{\theta _{w1}} - \theta_1  - {\bf{k}}\cdot{\bf{X}_1}} \right).
\end{equation}
In the resonant region, the normalized oscillation frequency exactly equals the gyro frequency as $\omega  = B\left( {{{\bf{X}}_0}} \right)/{\varepsilon }$ at some position $\mathbf{X}_0$. The absolute value of the coefficient of RHS of Eq.(\ref{g73}) and (\ref{g74}) almost equal in the resonant region. The main difference is the sign of the coefficient.

\section{Utilizing a new method to find out resonant terms on gyrocenter frame}\label{sec10}

In this section, we present a modified application of CLSLP, which in principle can find out all resonant terms in gyrocenter frame. The Lagrangian 1-form for test charged particle including the perturbation of magnetic vector potential and electrostatic potential is written as
\begin{eqnarray}\label{e1}
\gamma  =&& \left( {{\bf{A}}\left( {\bf{x}} \right) + {{\bf{A}}_w }\left( {{\bf{x}},t} \right)} \right)\cdot d{\bf{x}} + \varepsilon {\bf{v}}\cdot d{\bf{x}} \nonumber \\
&& - \left( {\varepsilon \frac{{{{\bf{v}}^2}}}{2} + \phi \left( {{\bf{x}},t} \right)} \right)dt,
\end{eqnarray}
As for the perturbation, the fourier branch of frequency $\omega$ and the wave vector $\bf{k}$ for each component of $\mathbf{A}_w(\mathbf{x},t)$ and $\phi(\mathbf{x},t)$ can be written as the superposition of two Fourier basis $\exp(i\omega t-i\mathbf{k}\cdot \mathbf{x})$ and $\exp(-i\omega t+i\mathbf{k}\cdot \mathbf{x})$, e.g., $\cos \left( {\omega t - {\bf{k}}\cdot{\bf{x}}} \right) = \left( {\exp \left( {i\omega t - i{\bf{k}}\cdot{\bf{x}}} \right) + \exp \left( { - i\omega t + i{\bf{k}}\cdot{\bf{x}}} \right)} \right)/2$. The former basis is recorded as $\Psi_-$, whilst $\Psi_+$ for the latter one. Then,  we use $\mathbf{A}_{\omega +}\Psi_+$ to denote the vector $(A_{w1+}\Psi_+\mathbf{e}_1, A_{w2+}\Psi_+\mathbf{e}_2, A_{w3+}\Psi_+\mathbf{e}_3)$, where $A_{wi+}$ for $i\in \{1,2,3\}$ is the amplitude for the basis $\Psi_+ \mathbf{e}_i$ . This notation is also applied to $\mathbf{A}_{w-}\Psi_{-}$ and $\phi_{w\pm}\Psi_{\pm}$. Then, we could get  $\mathbf{A}_{w}=\mathbf{A}_{w+}\Psi_{+}+\mathbf{A}_{w-}\Psi_{-}$ and $\phi_{w}=\phi_{w+}\Psi_{+}+\phi_{w-}\Psi_{-}$, respectively. The following analysis focuses on the single mode of $(\omega, \mathbf{k})$.

The following identities will be used:
\begin{equation}\label{ee1}
\nabla ({{\bf{A}}_{w \pm }}{\Psi _ \pm },{\phi _ \pm }{\Psi _ \pm }) =  \pm i{\bf{k}}({{\bf{A}}_{w \pm }}{\Psi _ \pm },{\phi _ \pm }{\Psi _ \pm }),
\end{equation}
\begin{equation}\label{ee2}
\nabla \times (\mathbf{A}_{w\pm}\Psi_{\pm})=\pm i \mathbf{k} \times {\mathbf{A}}_{w\pm}\Psi_{\pm},
\end{equation}
\begin{equation}\label{ee3}
\mathbf{\rho _0}\cdot{\bf{k}} = {\rho _0}{k_ \bot }\cos \alpha,
\end{equation}
where $\alpha$ is the angle between direction of $\mathbf{\rho}$ and $\mathbf{k}$, and equals the sum of $\theta$ and an constant angle,
\begin{equation}\label{ee4}
{\bm{\rho} _0}\cdot{{\bf{A}}_{w \pm }} = {\rho _0}\left( \begin{array}{l}
\frac{{{{\bf{e}}_1}}}{2}\left( {{e^{i\theta }} + {e^{ - i\theta }}} \right)\\
 - \frac{{{{\bf{e}}_2}}}{{2i}}\left( {\left( {{e^{i\theta }} - {e^{ - i\theta }}} \right)} \right)
\end{array} \right)\cdot{{\bf{A}}_{w \pm }}{\Psi _ \pm },
\end{equation}
with $\bm{\rho} _0 = \rho_0\left( { - {{\bf{e}}_1}\cos \theta  + {{\bf{e}}_2}\sin \theta } \right)$ and $\rho_0=\sqrt {\frac{{2\mu }}{{B\left( {\bf{x}} \right)}}} $.
The normalized Faraday's law is
\begin{equation}\label{ee3}
{\bf{E}}_{\pm} =  - {\partial _t}({{\bf{A}}_{w \pm }}\Psi_{\pm} )=  \pm i\omega {{\bf{A}}_{\omega  \pm }}\Psi_{\pm}.
\end{equation}

The original velocity vector $\bf{v}$ can be decomposed as three parts $v_\parallel \mathbf{b}$, $v_\perp \mathbf{\hat{v}}_\perp$ and $v_d\hat{\mathbf{v}}'_\perp$ with ${\widehat {\bf{v}}_ \bot } \equiv  {{{\bf{e}}_1}\sin \theta  + {{\bf{e}}_2}\cos \theta }$ and $\hat{\mathbf{v}}'_\perp$ perpendicular to $\mathbf{b}$ but independent on gyrophase.  $v_\parallel \mathbf{b}$ and $v_\perp \mathbf{\hat{v}}_\perp$ can be reexpressed in cylindrical coordinates by three components $(u,\mu,\theta)$ with  $u_1=\mathbf{v}\cdot \mathbf{b}$, $\mu_1=v_\perp^2/2B(\mathbf{x})$. $v_d\hat{\mathbf{v}}'_\perp$ is the drift velocity perpendicular to the direction of equilibrium magnetic field.
It is constituted by the drift velocity caused by the gradient and curvature of the magnetic field and the drift velocity caused by the electric field. The value of $v_d$ is usually much smaller than the other two velocities. So this drift velocity only appears in high order  and is ignored in our model.  Based on this approximation, $\gamma$ becomes
\begin{eqnarray}\label{e2}
\gamma  =&& \left( \begin{array}{l}
{{\bf{A}}}\left( {\bf{x}} \right) +{{\bf{A}}_{w\pm}}\Psi_{\pm} + {\varepsilon}{u_1}{\bf{b}}  \nonumber \\
 + {\varepsilon }\sqrt {2B({\bf{x}}){\mu _1}} {\widehat {\bf{v}}_ \bot }
\end{array} \right)\cdot d{\bf{x}}  \nonumber \\
&& - \left( {\varepsilon }{\frac{{u_1^2}}{2} + {\varepsilon }{\mu _1}B({\bf{x}})+\phi_{\pm}\Psi_{\pm}} \right)dt.
\end{eqnarray}
Since $\Psi_+$ and $\Psi_{-}$ are linearly independent from each other, their effects on the motion of particles are independently solved in Eq.(\ref{e2}). Eq.(\ref{e2}) can be separated into four parts
\begin{equation}\label{e3}
{\gamma _0} = {{\bf{A}}}\left( {\bf{x}} \right)\cdot d{\bf{x}},
\end{equation}
\begin{eqnarray}\label{e4}
{\gamma _1} =&& \left( {{u_1}{\bf{b}} + \sqrt {2B({\bf{x}}){\mu _1}} {{\widehat {\bf{v}}}_ \bot }} \right)\cdot d{\bf{x}} \nonumber \\
&& - \left( {\frac{{u_1^2}}{2} + {\mu _1}B({\bf{x}})} \right)dt,
\end{eqnarray}
\begin{equation}\label{e5}
{\gamma _{0w} } = {{\bf{A}}_{w\pm}}\Psi_{\pm}\cdot d{\bf{x}}.
\end{equation}
\begin{equation}\label{e6}
{\gamma _{0\sigma }} =  - \phi_{\pm} \Psi_{\pm}dt.
\end{equation}
It's assumed that there exists six independent parameters in this system. One is $\varepsilon$. The other ones come from the following order assumption: $O\left( {|\nabla  \times {{\bf{A}}_{w\pm}}|} \right) = O\left( {{\varepsilon _1}} \right)$, denoting the order of the amplitude of the perturbed magnetic field; $O\left( {|\partial {{\bf{A}}_{w\pm}}/\partial t|} \right) = O\left( {{\varepsilon _2}} \right)$, denoting the order of the amplitude of the inductive electric field; $O\left( {{\partial _ \bot }\ln |{{\bf{A}}_{w\pm}}|} \right) = O\left( {{\varepsilon _3^{-1}}} \right)$, denoting the order of the spatial gradient length of the perturbed magnetic vector potential; $O\left( {|\nabla \phi_{\pm} \left( {{\bf{x}},t} \right)|} \right) = O\left( {{\varepsilon _4^{-1}}} \right)$, denoting the order of the amplitude of the electrostatic potential; $O\left( {{\partial _ \bot }\ln |\phi_{\pm} ({\bf{x}},t)|} \right) = O\left( {{\varepsilon _5}} \right)$, denoting the order of the spatial gradient length of the electrostatic potential. All the orders are basically general assumption. In real physical systems, these six parameters may be related to each other by various physical mechanisms. As for its application in a real physical system,
it's only needed to replace each parameter by its real value. Then, the following expanding is applied to this perturbation problem.

We utilize the following Cary-Littlejohn single-parameter Lagrangian 1-form\cite{1983cary}
\begin{equation}\label{e7}
\Gamma  = \exp \left( { - \varepsilon {L_{{\bf{g}}_1^{\bf{X}}}}} \right)\gamma \left( {\bf{Z}} \right) + dS,
\end{equation}
which is transformed from the old one by the pullback transform.
To use this formula, a assumption is made that before and after coordinate transform, the time and particle's energy don't change. Hence, generators like $g^\mu,g^U,g^{\theta}, g^t$ are not introduced. The corresponding coordinate transform is ${\bf{x}} = \exp \left( { - {\bf{g}}_1^{\bf{X}}\cdot\nabla } \right){\bf{X}}$, ${\mu _1} = \mu $, ${u_1} = U$, and $\theta$ doesn't change.
Different from Ref.(\onlinecite{1983cary}), the gauge function is not used as a mean to cancel those unwanted terms. Eq.(\ref{e7}) can be expanded order by order according to the six parameters. To calculate the Lie derivative during the expanding, the following rules will be adopted
\begin{eqnarray}\label{m1}
&& {L_{{\bf{g}}_1^{\bf{x}}}}\left( {{\bf{f}}({\bf{Z}})\cdot d{\bf{X}}} \right) =  - {\bf{g}}_1^{\bf{x}} \times \nabla  \times {\bf{f}}\left( {\bf{Z}} \right)\cdot d{\bf{X}}  \nonumber \\
&& - {\bf{g}}_1^{\bf{x}}\cdot\left( {{\partial _t}{\bf{f}}({\bf{Z}})dt + {\partial _\theta }{\bf{f}}({\bf{Z}})d\theta  + {\partial _\mu }{\bf{f}}({\bf{Z}})d\mu } \right) + dS,
\end{eqnarray}
and
\begin{equation}\label{m2}
{L_{{\bf{g}}_1^{\bf{x}}}}\left( {h({\bf{Z}})dt} \right) = {\bf{g}}_1^{\bf{x}}\cdot\nabla h\left( {\bf{Z}} \right)dt+dS,
\end{equation}
where $\mathbf{f}(\bf{Z})$ and $h(\bf{Z})$ are arbitrary vector and scalar function of $\bf{Z}$, respectively.

Among the expanding terms, the following terms are recorded as $\Gamma_0$
\begin{eqnarray}\label{e8}
{\Gamma _0} =&& \left( {{\bf{A}}\left( {\bf{X}} \right) + {{\bf{A}}_{w\pm}}\Psi_{\pm}} \right)\cdot d{\bf{X}} \nonumber \\
 && - \phi_{\pm}\Psi_{\pm}dt.
\end{eqnarray}
In the 1-form, the terms $\mathbf{A}_{w\pm}\Psi_{\pm}$ and $\phi_{\pm}\Psi_{\pm}$ can be resonant terms if resonant condition $\omega-\mathbf{k}\cdot \dot{\mathbf{X}}\approx 0$ is satisfied. They represent the Landau resonance.

The term of order $O(\varepsilon)$ is recorded as
\begin{equation}\label{e9}
{\Gamma _{100000}} =\varepsilon U{\bf{b}}\cdot d{\bf{X}} - \varepsilon \left( {\mu B\left( {\bf{X}} \right) + \frac{{{U^2}}}{2}} \right)dt.
\end{equation}
Here, the subscripts of $\Gamma_{m_0 m_1 m_2 m_3 m_4 m_5}$ represent that $\Gamma_{m_0 m_1 m_2 m_3 m_4 m_5}$ is of order $O(\varepsilon^{m_0} \varepsilon_1^{m_1} \varepsilon_2^{m_2} \varepsilon_3^{m_3} \varepsilon_4^{m_4} \varepsilon_5^{m_5})$. It's well-known that to reach Eq.(\ref{e9}), the formula ${\bf{g}}_1^{\bf{X}} =  - {\rho _0}\left( { - {{\bf{e}}_1}\cos \theta  + {{\bf{e}}_2}\sin \theta } \right)$ is adopted with ${\rho _0} = \sqrt {\frac{{2\mu }}{{B\left( {\bf{X}} \right)}}}$.

The 1-form of order $O(\varepsilon \varepsilon_1)$ is
\begin{equation}\label{e22}
\begin{array}{l}
{\Gamma _{110000}} = \varepsilon \left[ {{\bf{g}}_1^{\bf{X}} \times \nabla  \times {{\bf{A}}_{w \pm }}{\Psi _ \pm }} \right]\cdot d{\bf{X}}\\
 \approx  \pm i\varepsilon \left[ {{\bf{g}}_1^{\bf{X}} \times \left( {{\bf{k}} \times {{\bf{A}}_{w \pm }}{\Psi _ \pm }} \right)} \right]\cdot d{\bf{X}}\\
 =  \pm i\varepsilon \left[ {\begin{array}{*{20}{l}}
{\frac{{{\rho _0}}}{2}\left( {\begin{array}{*{20}{l}}
{{{\bf{e}}_1}\left( {{e^{i\theta }} + {e^{ - i\theta }}} \right)}\\
{ - {{\bf{e}}_2}\frac{{\left( {{e^{i\theta }} - {e^{ - i\theta }}} \right)}}{i}}
\end{array}} \right)\cdot{{\bf{A}}_{w \pm }}{\Psi _ \pm }{\bf{k}}}\\
{ - \frac{{{\rho _0}}}{2}{k_ \bot }{{\bf{A}}_{w \pm }}{\Psi _ \pm }\left( {{e^{i\alpha }} + {e^{ - i\alpha }}} \right)}
\end{array}} \right]\cdot d{\bf{X}}.
\end{array}
\end{equation}
In the last equality of Eq.(\ref{e22}), Eqs.(\ref{ee3}) and (\ref{ee4}) are applied.
Eq.(\ref{e22}) includes resonant terms when resonant conditions  $\omega \pm \Omega  - {\bf{k}}\cdot{\dot{\mathbf{X}}}\approx 0$ are satisfied.

The 1-form of order $O(\varepsilon \varepsilon_2)$ is
\begin{equation}\label{e11}
\begin{array}{l}
{\Gamma _{101000}} = \varepsilon {\bf{g}}_1^{\bf{X}}\cdot{\partial _t}{{\bf{A}}_{w \pm }\Psi_{\pm}}dt\\
 =  \pm i\varepsilon \omega {\rho _0}\left( {\begin{array}{*{20}{l}}
{{{\bf{e}}_1}\left( {{e^{i\theta }} + {e^{ - i\theta }}} \right)/2}\\
{ - {{\bf{e}}_2}\left( {{e^{i\theta }} - {e^{ - i\theta }}} \right)/2i}
\end{array}} \right)\cdot{{\bf{A}}_{w \pm }\Psi_{\pm}}dt.
\end{array}
\end{equation}
Eq.(\ref{e11}) includes resonant terms when resonant conditions $\omega \pm \Omega-\mathbf{k}\cdot \dot{\mathbf{X}}\approx 0$ are satisfied.

The 1-form of order $O(\varepsilon^2 \varepsilon_1 \varepsilon_3)$ is
\begin{equation}\label{e12}
\begin{array}{*{20}{l}}
{{\Gamma _{210100}} = {\varepsilon ^2}\left[ {{\bf{g}}_1^{\bf{X}} \times \nabla  \times \left( {{\bf{g}}_1^{\bf{X}} \times \nabla  \times {{\bf{A}}_{w \pm }}{\Psi _ \pm }} \right)} \right]\cdot d{\bf{X}}}\\
{ \approx  - {\varepsilon ^2}\left[ {{\bf{g}}_1^{\bf{X}} \times \left[ {{\bf{k}} \times \left( {{\bf{g}}_1^{\bf{X}} \times \left( {{\bf{k}} \times {{\bf{A}}_{w \pm }}} \right)} \right)} \right]} \right]{\Psi _ \pm }\cdot d{\bf{X}}}\\
\begin{array}{l}
 = {\varepsilon ^2}\left( {{\bf{g}}_1^{\bf{X}}\cdot{\bf{k}}} \right)\left( {{\bf{g}}_1^{\bf{X}}\cdot{{\bf{A}}_{w \pm }}} \right){\Psi _ \pm }{\bf{k}}\cdot d{\bf{X}}\\
 - {\varepsilon ^2}{\left( {{\bf{g}}_1^{\bf{X}}\cdot{\bf{k}}} \right)^2}{{\bf{A}}_{w \pm }}{\Psi _ \pm }\cdot d{\bf{X}}
\end{array}\\
{ = \frac{{{\varepsilon ^2}\rho _0^2{k_ \bot }}}{4}\left( {{e^{i\alpha }} + {e^{ - i\alpha }}} \right)\left( \begin{array}{l}
{{\bf{e}}_1}\left( {{e^{i\theta }} + {e^{ - i\theta }}} \right)\\
 - \frac{{{{\bf{e}}_2}\left( {{e^{i\theta }} - {e^{ - i\theta }}} \right)}}{i}
\end{array} \right)}\\
{\cdot{{\bf{A}}_{w \pm }}{\Psi _ \pm }{\bf{k}}\cdot d{\bf{X}} - \frac{{{\varepsilon ^2}k_ \bot ^2\rho _0^2}}{2}\left( \begin{array}{l}
1 + \frac{{{e^{ - i2\alpha }}}}{2}\\
 + \frac{{{e^{i2\alpha }}}}{2}
\end{array} \right){{\bf{A}}_{w \pm }}{\Psi _ \pm }\cdot d{\bf{X}}}
\end{array}
\end{equation}
Eq.(\ref{e12}) includes resonant terms when resonant conditions $\omega  \pm 2\Omega  - {\bf{k}}\cdot\dot{{\mathbf{X}}}\approx 0$ are satisfied.

The 1-form of order $O(\varepsilon^2\varepsilon_1\varepsilon_2)$ is
\begin{equation}\label{e13}
\begin{array}{l}
{\Gamma _{211000}} = {\varepsilon ^2}\left[ {{\bf{g}}_1^{\bf{X}}\cdot{\partial _t}\left( {{\bf{g}}_1^{\bf{X}} \times \nabla  \times {{\bf{A}}_{w \pm }}\Psi_{\pm}} \right)} \right]dt\\
 \approx {\varepsilon ^2}\omega \left[ {{\bf{g}}_1^{\bf{X}}\cdot\left( {\begin{array}{*{20}{l}}
{\left( {{\bf{g}}_1^{\bf{X}}\cdot{{\bf{A}}_{w \pm }\Psi_{\pm}}} \right){\bf{k}}}\\
{ - \left( {{\bf{g}}_1^{\bf{X}}\cdot{\bf{k}}} \right){{\bf{A}}_{w \pm }}\Psi_{\pm}}
\end{array}} \right)} \right]dt = 0.
\end{array}
\end{equation}

The 1-form of order $O(\varepsilon^2 \varepsilon_2 \varepsilon_3)$ is
\begin{equation}\label{e14}
\begin{array}{*{20}{l}}
{{\Gamma _{201100}} =  - {\varepsilon ^2}{\bf{g}}_1^{\bf{X}}\cdot\nabla \left( {{\bf{g}}_1^{\bf{X}}\cdot{\partial _t}{{\bf{A}}_{w \pm }}{\Psi _ \pm }} \right)dt}\\
{ \approx {\varepsilon ^2}\omega \left( {{\bf{g}}_1^{\bf{X}}\cdot{\bf{k}}} \right)\left( {{\bf{g}}_1^{\bf{X}}\cdot{{\bf{A}}_{w \pm }}{\Psi _ \pm }} \right)dt}\\
{\begin{array}{*{20}{l}}
{ = \frac{1}{4}{\varepsilon ^2}\omega \rho _0^2{k_ \bot }\left( {{e^{i\alpha }} + {e^{ - i\alpha }}} \right)}\\
{ \times \left( \begin{array}{l}
{{\bf{e}}_1}\left( {{e^{i\theta }} + {e^{ - i\theta }}} \right)\\
 - {{\bf{e}}_2}\left( {{e^{i\theta }} - {e^{ - i\theta }}} \right)/i
\end{array} \right)\cdot{{\bf{A}}_{w \pm }}{\Psi _ \pm }dt}
\end{array}}
\end{array}
\end{equation}
There are resonant terms in Eq.(\ref{e14}) when resonant conditions $\omega  \pm 2\Omega  - {\bf{k}}\cdot\dot{{\mathbf{X}}}\approx 0$ are satisfied.

The 1-form of order $O(\varepsilon \varepsilon_{4})$ is
\begin{equation}\label{e15}
\begin{array}{*{20}{l}}
{{\Gamma _{100010}} = \varepsilon {\bf{g}}_1^{\bf{X}}\cdot\nabla {\phi _ \pm }{\Psi _ \pm }dt \approx  \pm i\varepsilon \left( {{\bf{g}}_1^{\bf{X}}\cdot{\bf{k}}} \right){\phi _ \pm }{\Psi _ \pm }}\\
{ =  \mp i\frac{{\varepsilon {\rho _0}{k_ \bot }}}{2}\left( {{e^{i\alpha }} + {e^{ - i\alpha }}} \right){\phi _ \pm }{\Psi _ \pm }.}
\end{array}
\end{equation}
Its resonant conditions are $\omega  \pm \Omega  - {\bf{k}}\cdot\dot{{\mathbf{X}}}\approx 0$.

The 1-form of order $O(\varepsilon^2 \varepsilon_4 \varepsilon_5)$ is
\begin{equation}\label{e16}
\begin{array}{*{20}{l}}
{{\Gamma _{200011}} = {\varepsilon ^2}{{\left( {{\bf{g}}_1^{\bf{X}}\cdot\nabla } \right)}^2}{\phi _ \pm }{\Psi _ \pm }dt}\\
{ \approx  - {\varepsilon ^2}{{\left( {{\bf{g}}_1^{\bf{X}}\cdot{\bf{k}}} \right)}^2}{\phi _ \pm }{\Psi _ \pm }dt}\\
{ = \frac{1}{2}{\varepsilon ^2}k_ \bot ^2\rho _0^2\left( {1 + \frac{{{e^{ - i2\alpha }}}}{2} + \frac{{{e^{i2\alpha }}}}{2}} \right){\phi _ \pm }{\Psi _ \pm }dt.}
\end{array}
\end{equation}
Its resonant conditions are $\omega  \pm 2\Omega  - {\bf{k}}\cdot\dot{{\mathbf{X}}}\approx 0$.

The 1-form of order $O(\varepsilon^2)$ is recorded as
\begin{eqnarray}\label{e35}
{\Gamma _{200000}} &=& \frac{{{\varepsilon ^2}}}{2}{L_{{\bf{g}}_1^{\bf{x}}}}{\gamma _0}\left( {\bf{Z}} \right) - {L_{{\bf{g}}_1^x}}{\gamma _1}\left( {\bf{Z}} \right) + dS  \nonumber \\
 &=& \mu d\theta  + {\Gamma _{2f}}({\bf{Z}}),
\end{eqnarray}
where ${\Gamma _{2f}}({\bf{Z}})$ represents the fast part included by $\Gamma_{200000}$.  $\Gamma_{200000}$ includes a term $\varepsilon^2 \mu d\theta$, which is like a action and angle term. This term helps to determine the evolution of $\mu$.

The expanding is stopped here.  The higher order terms can be treated with the same procedure. The foregoing expanding is based on the assumption that all the parameters are independent from each other. As for application, each parameter is given a specific number. If two parameters are of the same order, the respective expanding for each parameter are combined together as the expanding upon this parameter. For example, if $O(\varepsilon_3)=O(\varepsilon_5)$, the sum of the expanding over $\varepsilon_3$ and $\varepsilon_5$ are treated as the expanding over this parameter.

We only consider up to the second harmonic resonance, since experiments in tokamak plasma usually use the fundamental frequency resonance for the minority heating and the second harmonic frequency resonance for the majority heating\cite{wesson2004}. So even the 1-form responsible for the third harmonic resonance is of order lower than $O(\varepsilon^2)$, it's neglected in this paper.  By summing all the expanding terms together,  a new 1-form is obtained
\begin{eqnarray}\label{e17}
{\Gamma _0} =& \left( {{\bf{A}}\left( {\bf{X}} \right) + {{\bf{A}}_1}\left( {\bf{Z}} \right) + \varepsilon U{\bf{b}}} \right)\cdot d{\bf{X}}+\varepsilon^2 \mu d\theta \nonumber \\
& - \left[ {\varepsilon \mu B\left( {\bf{X}} \right) + \varepsilon \frac{{{U^2}}}{2} + {\phi _1}\left( {\bf{Z}} \right)} \right]dt,
\end{eqnarray}
where $\mathbf{A}_1(\mathbf{Z})$ and $\phi_1(\mathbf{Z})$ denote the sum of those expanding terms associated with the perturbations $\mathbf{A}_{w}$ and $\phi_w$ in $\mathbf{X}$ and $t$ components up to the second harmonic resonance, respectively.
The trajectory equations derived from Eq.(\ref{e17}) are
\begin{equation}\label{e18}
\mathop {\bf{X}}\limits^. {\rm{ = }}\frac{{U{{\bf{B}}^*} + {\bf{b}} \times \nabla \left( {{H_0} + {\phi _1}\left( {\bf{Z}} \right)} \right) + \partial {{\bf{A}}_1}({\bf{Z}})/\partial t \times {\bf{b}}}}{{{\bf{b}}\cdot{{\bf{B}}^*}}},
\end{equation}
\begin{equation}\label{e19}
\begin{array}{l}
\dot U = \frac{{ - \left( {{{\bf{B}}^*}\cdot\nabla \left( {{H_0} + {\phi _1}\left( {\bf{Z}} \right)} \right)} \right) + {\bf{b}}\cdot\left[ {\left( {\partial {{\bf{A}}_1}\left( {\bf{Z}} \right)/\partial t \times {\bf{b}}} \right) \times {{\bf{B}}^*}} \right]}}{{\varepsilon {\bf{b}}\cdot{\bf{B}}{\rm{*}}}}\\
 - {\bf{b}}\cdot\frac{{\partial {{\bf{A}}_1}({\bf{Z}})}}{{\varepsilon \partial t}},
\end{array}
\end{equation}
\begin{equation}\label{e20}
\dot \theta  = \frac{{B\left( {\bf{X}} \right)}}{\varepsilon } + \frac{1}{{{\varepsilon ^2}}}\frac{{\partial {\phi _1}\left( {\bf{Z}} \right)}}{{\partial \mu }} - \frac{1}{{{\varepsilon ^2}}}\frac{{\partial {{\bf{A}}_1}\left( {\bf{Z}} \right)\cdot\mathop {\bf{X}}\limits^. }}{{\partial \mu }},
\end{equation}
\begin{eqnarray}\label{e21}
\dot \mu  &=& \frac{1}{{{\varepsilon ^2}}}\frac{\partial }{{\partial \theta }}{{\bf{A}}_1}\left( {\bf{Z}} \right)\cdot\mathop {\bf{X}}\limits^.  - \frac{1}{{{\varepsilon ^2}}}\frac{\partial }{{\partial \theta }}{\phi _1}\left( {\bf{Z}} \right) \nonumber \\
& \approx & \frac{U}{{{\varepsilon ^2}}}\frac{\partial }{{\partial \theta }}{A_{1\parallel }}\left( {\bf{Z}} \right) - \frac{1}{{{\varepsilon ^2}}}\frac{\partial }{{\partial \theta }}{\phi _1}\left( {\bf{Z}} \right).
\end{eqnarray}
The cyclotron frequency $\frac{B(\mathbf{X})}{\varepsilon}$ is positive for particles of positive charge, and is negative for particles of negative charge.

In App.(\ref{app1}), for each resonant condition, all resonant terms are isolated out from $\mathbf{A}_1(\bf{Z})$ and $\phi_1(\bf{Z})$ to form the slow-dynamic terms $\mathbf{A}_{1s}$ and $\phi_{1s}$. By substituting these terms back into Eqs.(\ref{e18}-\ref{e21}), evolution equations for the slow parts of $\{\mathbf{X}_s,U_s,\mu_s\}$ as well as $\theta$ are derived. Here, subscript ``s" is used as the symbol of the slow part.

One important character revealed by the trajectory equation of $\theta$ is that the frequency of $\theta$ is shifted from the normalized cyclotron frequency $\Omega_0=B(\bf{X})/\varepsilon$ under the driving of resonant wave.

\section{An example for fundamental-frequency cyclotron resonance}\label{sec13}

In this section, a simple example of driving the proton by the circularly polarized wave is given to show the resonant effect and the shift of frequency described by the slow-dynamic trajectory equations. The spatially homogeneous magnetic field configuration is chosen to be directed in $z$ direction with the  amplitude $1T$. The wave is only of the wave vector parallel to $\mathbf{e}_z$. It's circularly polarized and it's expressed by the formula
\begin{equation}\label{e36}
\begin{array}{l}
{{\bf{A}}_w} = 2\psi \left( {{{\bf{e}}_1}\cos \left( {\omega t - {k_\parallel }z} \right) - {{\bf{e}}_2}\sin \left( {\omega t - {k_\parallel }z} \right)} \right)\\
 = {{\bf{A}}_ - }\Psi_- + {{\bf{A}}_ + }\Psi_{+},
\end{array}
\end{equation}
with ${{\bf{A}}_ - } = \psi \left( {{{\bf{e}}_1} - \frac{{{{\bf{e}}_2}}}{i}} \right)$ and ${{\bf{A}}_ +} = \psi \left( {{{\bf{e}}_1} + \frac{{{{\bf{e}}_2}}}{i}} \right)$.
The quantities used for the normalization are $B_0=1T,v_t=10^{5}m/s,L_0=1m$. The normalization scheme is given before. With these quantities, it's derived that $\varepsilon=10^{-3}$. The normalized values of $\psi$ and $k_\parallel$ are given to be $\psi=10^{-4}$ and $k_\parallel=40.0$.  The frequency $\omega$ will be determined later. Substituting Eq.(\ref{e36}) back into Eqs.(\ref{e28},\ref{e29}), it's derived that
\begin{equation}\label{e37}
{{\bf{A}}_{1s}} = 2\psi \varepsilon k_\parallel {\rho _0}\sin \left( {\omega t - \theta  - {k_\parallel }z}+\xi \right),
\end{equation}
\begin{equation}\label{e38}
{\phi _{1s}} =  - 2\varepsilon \omega {\rho _0}\psi \sin \left( {\omega t - \theta  - {k_\parallel }z} +\xi \right),
\end{equation}
where $\xi$ is an initial phase and can be an arbitrary angle.
By substituting Eqs.(\ref{e37}) and (\ref{e38}) back into Eqs.(\ref{e18})-(\ref{e21}), the slow-dynamic equations are derived
\begin{equation}\label{e39}
\mathop {\bf{X}}\limits^.  = U,
\end{equation}
\begin{equation}\label{e40}
\dot U =  - 4\psi {\rho _0}{k_\parallel }\omega \cos \left( {\omega t - \theta  - {k_\parallel }z + \xi } \right),
\end{equation}
\begin{equation}\label{e41}
\dot \theta  = \frac{1}{\varepsilon }- \frac{{2\omega \psi }}{\varepsilon }\frac{1}{{\sqrt {2\mu } }}\sin \left( {\omega t - \theta  - {k_\parallel }z + \xi } \right),
\end{equation}
\begin{equation}\label{e31}
\begin{array}{l}
\dot \mu  = \frac{{ - 2U}}{\varepsilon }\psi {k_\parallel }{\rho _0}\left( {\omega t - \theta  - {k_\parallel }z + \xi } \right)\\
 - \frac{2}{\varepsilon }\omega \psi {\rho _0}\cos \left( {\omega t - \theta  - {k_\parallel }z + \xi } \right),
\end{array}
\end{equation}
where the subscript $``s"$ is deleted for all variables for convenience.
Here, although the phase ${\omega t - \theta  - {k_\parallel }z + \xi }$ depends on $\theta$, it's not fast changing anymore, due to the resonance.
One important point of this group of trajectory equations is that the frequency of $\theta$ is not the cyclotron frequency $\Omega_0=1/\varepsilon$ anymore, but is shifted by the wave driving. The shifting part is inversely proportional to $\sqrt{2\mu}$,  while proportional to the amplitude of the driving electric field. Since $\mu$ changes by the wave driving with changing rate determined by Eq.(\ref{e31}), it's impossible to choose a special frequency to satisfy the accurate cyclotron resonance for long time even ignoring the frequency mismatch caused by the changing of parallel velocity as Eq.(\ref{e40}) shows. A detailed investigation of the shift of the cyclotron resonant frequency is given in future work. Here, we present a simple example to show the shift of cyclotron resonant frequency and its influence on the resonant driving of magnetic moment.

The initial conditions for the test ion are chosen to be $U=0.5,\mu=0.125,X=Y=Z=0$. The initial phase $\xi$ is chosen to be $1.3\pi$. Enlightened by Eq.(\ref{e41}), we choose two waves of different frequency to make comparison. The first frequency is ${\omega _1} = \Omega_0  + 20 + 0.64 \times \Omega_0 \psi /\varepsilon $, whilst the second one is ${\omega _2} = \Omega_0  + 20$. The difference between $\omega_1$ and $\omega_2$ is about $1/10$ of $\Omega_0$. The quantity $20$ is introduced to cancel the initial phase frequency $Uk_\parallel$. The real amplitude of the electric field with normalized $\psi$ is about $10^4$V/m, which can be reached by currently applied ICRH heating \cite{1981hwang}.

\begin{figure}[htbp]
%\centering
\includegraphics[height=6cm,width=7cm]{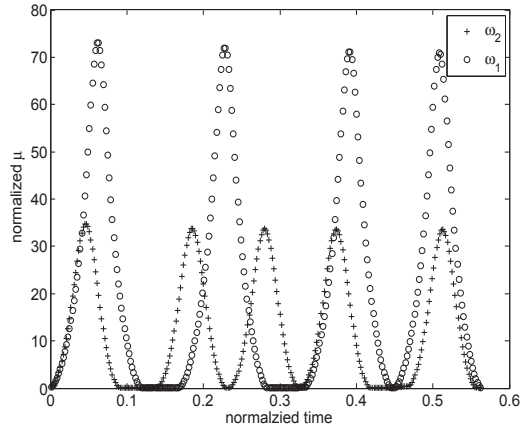}  %\centering
\caption{\label{frecom}  The comparison of the driving effect for the normalized $\mu$ by the waves of frequency $\omega_1$ and $\omega_2$, respectively. With a shifting frequency deviating from the cyclotron frequency, the wave of frequency $\omega_1$ has a much better driving effect due to a longer time of phase match, and can drive the ion to a much higher kinetic energy.}
\end{figure}

For numerical calculation, the time step is given to be $dt=10^{-3}$. Attention should be given to the numerical simulation for the dependence of the frequency of $\theta$ on $\mu$ , and to the composite phase as well as the positive property of the value of $\mu$. The numerical comparison between the driving by the two waves is given in Fig.(\ref{frecom}).
The important point shown by Fig.(\ref{frecom}) is that the wave of shifting resonance frequency $\omega_1$ is more effective to drive magnetic moment to a larger value, due to the longer time for the phase match between the driving electric field and the cyclotron motion of charged particle.  Another phenomenon in Fig.(\ref{frecom}) is that some wave troughs are wider than others. This is determined by the composite phase. For the wider trough, the changing rate of the composite phase at that time interval is slower to inverse the decelerating of the perpendicular velocity to the accelerating of perpendicular velocity. The inverse process is determined by Eq.(\ref{e31}).

%Noting the shifting frequency is about $1/10$ of the normalized $\Omega_0$ equaling $10^{-3}$ under the given simulating condition, it should be given notation.

\section{Summary and Discussion}\label{sec8}

This paper pointed out that the application of Lie transform method in GT generates some nonphysical terms, in particular, the nonphysical cyclotron resonant term. A modified edition is developed to generate the resonant terms for various resonant conditions up to the second harmonic resonance. By getting rid of the fast-dynamical terms from the trajectory equations, the rest equations are governed by slow dynamics including the resonant effects.  One important revelation of these trajectory equations is that the optimal driving frequency of the wave for the cyclotron resonance is not the original cyclotron frequency without the driving of the wave, but a frequency deviating from the original frequency.

One application of these slow-dynamic equations is to calculate straightforwardly the orbit of charged particle passing through the resonant layer driving by the resonant electromagnetic wave, and to observe the variation of energy and pitch angle.

%By the way, there in fact exists another problem in the modified cyclotron resonant GT as well as in  cyclotron resonant modern GT. As can be observed from Eq.(\ref{g36}), the cyclotron resonance origins from  term $\left( {\exp \left( {{\rho _0}\cdot{\nabla _{\bf{X}}}} \right){{\bf{A}}_w}\cdot\frac{{\partial {\rho _0}}}{{\partial \theta }} - {\varepsilon ^*}{g^\mu }} \right)d\theta $. This term can only give funta

\section{Acknowledgments}\label{sec9}

The author thanks the discussion with Prof. Jiquan Li, Prof. Yasuaki Kishimoto and Dr. Wu Feng. This work was completed at Uji campus of Kyoto University, Japan. This work is partially supported by Chinese Scholarship Council with  NO: 201504890004.

% and was partially supported by the scholarship of number CSC NO.201504890004.

\appendix

\section{Resonant terms for various resonant conditions}\label{app1}

Based on the identities $\mathbf{A}_{w}=\mathbf{A}_{w+}\Psi_{+}+\mathbf{A}_{w-}\Psi_{-}$ and $\phi_{w}=\phi_{w+}\Psi_{+}+\phi_{w-}\Psi_{-}$, the resonant terms for various resonant conditions are derived as follows.

\subsection{Resonant condition $\omega-\mathbf{k}\cdot \dot{\mathbf{X}}\approx 0$}\label{sec11.1}

The terms responsible for this resonant condition are $\mathbf{A}_{w\pm}\Psi_{\pm}$ and $\phi_{\pm}\Psi_{\pm}$. The slow-dynamics parts of $\mathbf{A}_{1}(\bf{Z})$ and $\phi_{1}(\bf{Z})$ in Eqs.(\ref{e18},\ref{e19}) are $\mathbf{A}_{s1}=\mathbf{A}_{w+}\Psi_{+}+\mathbf{A}_{w-}\Psi_{-}$ and $\phi_{s1}=\phi_{+}\Psi_{+}+\phi_{-}\Psi_{-}$, respectively. Other terms included in $\mathbf{A}_{1}(\bf{Z})$ and $\phi_{1}(\bf{Z})$ are responsible for fast dynamics  and are separated away. The evolution equation for $\mu$ is $\dot{\mu}=0$.

\subsection{Resonant condition $\omega-\Omega-\mathbf{k}\cdot \dot{\mathbf{X}}\approx 0$}\label{sec11.2}

The terms satisfying this kind of resonant condition are of the phase like $\omega t - {\bf{k}}\cdot{\bf{X}} - \theta  + \xi$.

For this resonant condition , the resonant terms corresponding to the the slow-dynamic parts of $\mathbf{A}_{1}$ and $\phi_{1}$ are
\begin{eqnarray}\label{e28}
{{\bf{A}}_{1s}}\left( {\bf{Z}} \right) &&=  i\frac{\varepsilon {\rho _0} }{2}{e^{i\theta }}\left( {{{\bf{e}}_1} - \frac{{{{\bf{e}}_2}}}{i}} \right)\cdot{{\bf{A}}_{w + }}{\Psi _ + }{\bf{k}} \nonumber \\
 &&- i\frac{\varepsilon {\rho _0}}{2}{e^{ - i\theta }}\left( {{{\bf{e}}_1} + \frac{{{{\bf{e}}_2}}}{i}} \right)\cdot{{\bf{A}}_{w - }}{\Psi _ - }{\bf{k}}  \nonumber \\
 && - i\frac{\varepsilon {\rho _0} {k_ \bot }}{2}\left( {{{\bf{A}}_{w + }}{\Psi _ + }{e^{i\alpha }} - {{\bf{A}}_{w - }}{\Psi _ - }{e^{ - i\alpha }}} \right),
\end{eqnarray}
and
\begin{eqnarray}\label{e29}
{\phi _{1s}}\left( {\bf{Z}} \right) && \approx i\frac{\varepsilon \omega {\rho _0}}{2}{e^{i\theta }}\left( {{{\bf{e}}_1} - \frac{{{{\bf{e}}_2}}}{i}} \right)\cdot{{\bf{A}}_{w + }}{\Psi _ + } \nonumber \\
&& - i\frac{\varepsilon \omega {\rho _0}}{2}{e^{ - i\theta }}\left( {{{\bf{e}}_1} + \frac{{{{\bf{e}}_2}}}{i}} \right)\cdot{{\bf{A}}_{w - }}{\Psi _ - } \nonumber \\
&& - i\frac{{\varepsilon {\rho _0}{k_ \bot }}}{2}\left( {{e^{i\alpha }}{\phi _ + }{\Psi _ + } - {e^{ - i\alpha }}{\phi _ - }{\Psi _ - }} \right).
\end{eqnarray}

These terms can be responsible for the fundamental-frequency cyclotron resonance of particles of positive charge.

\subsection{Resonant condition $\omega+\Omega-\mathbf{k}\cdot \dot{\mathbf{X}}\approx 0$}\label{sec11.3}

The terms satisfying this kind of resonant condition are of the phase like $ \omega t - {\bf{k}}\cdot{\bf{X}} + \theta  + \xi$.

For this resonant condition, the resonant terms corresponding to the the slow-dynamic parts of $\mathbf{A}_{1}$ and $\phi_{1}$ are
\begin{eqnarray}\label{e30}
{{\bf{A}}_{1s}}\left( {\bf{Z}} \right) && = i\frac{\varepsilon {\rho _0}}{2}{e^{i\theta }}\left( {{{\bf{e}}_1} - \frac{{{{\bf{e}}_2}}}{i}} \right)\cdot{{\bf{A}}_{w - }}{\Psi _ - }{\bf{k}} \nonumber \\
&& - i\frac{\varepsilon {\rho _0}}{2}{e^{ - i\theta }}\left( {{{\bf{e}}_1} + \frac{{{{\bf{e}}_2}}}{i}} \right)\cdot{{\bf{A}}_{w + }}{\Psi _ + }{\bf{k}} \nonumber \\
&& - i\frac{\varepsilon {\rho _0}{k_ \bot }}{2}\left( {{{\bf{A}}_{w + }}{\Psi _ + }{e^{ - i\alpha }} - {{\bf{A}}_{w - }}{\Psi _ - }{e^{ + i\alpha }}} \right),
\end{eqnarray}
and
\begin{eqnarray}\label{e311}
{\phi _{1s}}\left( {\bf{Z}} \right)&& \approx i\frac{{\varepsilon \omega {\rho _0}}}{2}{e^{ - i\theta }}\left( {{{\bf{e}}_1} + \frac{{{{\bf{e}}_2}}}{i}} \right)\cdot{{\bf{A}}_{w + }}{\Psi _ + }\\
&& - i\frac{{\varepsilon \omega {\rho _0}}}{2}{e^{i\theta }}\left( {{{\bf{e}}_1} - \frac{{{{\bf{e}}_2}}}{i}} \right)\cdot{{\bf{A}}_{w - }}{\Psi _ - }\\
&& - i\frac{{\varepsilon {\rho _0}{k_ \bot }}}{2}\left( {{e^{ - i\alpha }}{\phi _ + }{\Psi _ + } - {e^{i\alpha }}{\phi _ - }{\Psi _ - }} \right).
\end{eqnarray}

For particles of negative charge, $\varepsilon<0$ stands, leading to the negative value of $\Omega_0$ equaling $\frac{B(\mathbf{X})}{\varepsilon}$. So these terms are responsible for the fundamental-frequency cyclotron resonance of particles of negative charges.

\subsection{Resonant condition $\omega-2\Omega-\mathbf{k}\cdot \dot{\mathbf{X}}\approx 0$}\label{sec11.4}

The terms satisfying this kind of resonant condition are of the phase like $ \omega t - {\bf{k}}\cdot{\bf{X}} - 2\theta  + \xi$.

For this resonant condition, the resonant terms corresponding to the the slow-dynamic parts of $\mathbf{A}_{1}$ and $\phi_{1}$ are
\begin{eqnarray}\label{e32}
{{\bf{A}}_{1s}}&& = \frac{{{\varepsilon ^2}\rho _0^2{k_ \bot }}}{4}{e^{i\alpha  + i\theta }}\left( {{{\bf{e}}_1} - \frac{{{{\bf{e}}_2}}}{i}} \right)\cdot{{\bf{A}}_{w + }}{\Psi _ + }{\bf{k}} \nonumber \\
&& + \frac{{{\varepsilon ^2}\rho _0^2{k_ \bot }}}{4}{e^{ - i\alpha  - i\theta }}\left( {{{\bf{e}}_1} + \frac{{{{\bf{e}}_2}}}{i}} \right)\cdot{{\bf{A}}_{w - }}{\Psi _ - }{\bf{k}}\nonumber \\
&& - \frac{{{\varepsilon ^2}k_ \bot ^2\rho _0^2}}{4}\left( {{e^{i2\alpha }}{{\bf{A}}_{w + }}{\Psi _ + } + {e^{ - i2\alpha }}{{\bf{A}}_{w - }}{\Psi _ - }} \right),
\end{eqnarray}
and
\begin{eqnarray}\label{e33}
{\phi _{1s}}\left( {\bf{Z}} \right)  \approx && \frac{{\varepsilon ^2}\rho _0^2{k_ \bot }\omega}{4} \left[ \begin{array}{l}
{e^{i\alpha  + i\theta }}\left( {{{\bf{e}}_1} - \frac{{{{\bf{e}}_2}}}{i}} \right)\cdot{{\bf{A}}_{w + }}{\Psi _ + }  \nonumber \\
 + {e^{ - i\alpha  - i\theta }}\left( {{{\bf{e}}_1} + \frac{{{{\bf{e}}_2}}}{i}} \right)\cdot{{\bf{A}}_{w - }}{\Psi _ - }
\end{array} \right]  \nonumber \\
&& + \frac{{\varepsilon ^2}k_ \bot ^2\rho _0^2}{4}\left( {{e^{ - i2\alpha }}{\phi _ - }{\Psi _ - } + {e^{i2\alpha }}{\phi _ + }{\Psi _ + }} \right).
\end{eqnarray}

These terms can be responsible for the second harmonic cyclotron resonance of particles of positive charge.

\subsection{Resonant condition $\omega+2\Omega-\mathbf{k}\cdot \dot{\mathbf{X}}\approx 0$}\label{sec11.5}

The terms satisfying this kind of resonant condition are of the phase like $\omega t - {\bf{k}}\cdot{\bf{X}} + 2\theta  + \xi$.

For this resonant condition, the resonant terms corresponding to the the slow-dynamic parts of $\mathbf{A}_{1}$ and $\phi_{1}$ are
\begin{eqnarray}\label{e32}
{{\bf{A}}_{1s}} &&= \frac{{{\varepsilon ^2}\rho _0^2{k_ \bot }}}{4}{e^{i\alpha  + i\theta }}\left( {{{\bf{e}}_1} - \frac{{{{\bf{e}}_2}}}{i}} \right)\cdot{{\bf{A}}_{w - }}{\Psi _ - }{\bf{k}} \nonumber \\
&& + \frac{{{\varepsilon ^2}\rho _0^2{k_ \bot }}}{4}{e^{ - i\alpha  - i\theta }}\left( {{{\bf{e}}_1} + \frac{{{{\bf{e}}_2}}}{i}} \right)\cdot{{\bf{A}}_{w + }}{\Psi _ + }{\bf{k}} \nonumber \\
&& - \frac{{{\varepsilon ^2}k_ \bot ^2\rho _0^2}}{4}\left( {{e^{i2\alpha }}{{\bf{A}}_{w - }}{\Psi _ - } + {e^{ - i2\alpha }}{{\bf{A}}_{w + }}{\Psi _ + }} \right),
\end{eqnarray}
and
\begin{eqnarray}\label{e33}
{\phi _{1s}} =&& \frac{{{\varepsilon ^2}\omega \rho _0^2{k_ \bot }}}{4}\left[ \begin{array}{l}
{e^{i\alpha  + i\theta }}\left( {{{\bf{e}}_1} - \frac{{{{\bf{e}}_2}}}{i}} \right)\cdot{{\bf{A}}_{w - }}{\Psi _ - } \\
{e^{ - i\alpha  - i\theta }}\left( {{{\bf{e}}_1} + \frac{{{{\bf{e}}_2}}}{i}} \right)\cdot{{\bf{A}}_{w + }}{\Psi _ + }
\end{array} \right] \nonumber \\
 &&+ \frac{{\varepsilon ^2}k_ \bot ^2\rho _0^2}{4}\left( {{e^{ - i2\alpha }}{\phi _ + }{\Psi _ + } + {e^{i2\alpha }}{\phi _ - }{\Psi _ - }} \right).
\end{eqnarray}

These terms can be responsible for the second harmonic cyclotron resonance of particles of negative charge.

% If in two-column mode, this environment will change to single-column format so that long equations can be displayed.
%\begin{widetext}
%$$\mbox{put long equation here}$$
%\end{widetext}

% Figures should be put into the text as floats.
% Use the graphics or graphicx packages (distributed with LaTeX2e).
% Put the label that you will use with \ref{} command in the braces of the \label{} command.

%\bibliographystyle{aipauth4-1}
%\bibliography{hfheat}

\begin{thebibliography}{29}%
\makeatletter
\providecommand \@ifxundefined [1]{%
 \@ifx{#1\undefined}
}%
\providecommand \@ifnum [1]{%
 \ifnum #1\expandafter \@firstoftwo
 \else \expandafter \@secondoftwo
 \fi
}%
\providecommand \@ifx [1]{%
 \ifx #1\expandafter \@firstoftwo
 \else \expandafter \@secondoftwo
 \fi
}%
\providecommand \natexlab [1]{#1}%
\providecommand \enquote  [1]{``#1''}%
\providecommand \bibnamefont  [1]{#1}%
\providecommand \bibfnamefont [1]{#1}%
\providecommand \citenamefont [1]{#1}%
\providecommand \href@noop [0]{\@secondoftwo}%
\providecommand \href [0]{\begingroup \@sanitize@url \@href}%
\providecommand \@href[1]{\@@startlink{#1}\@@href}%
\providecommand \@@href[1]{\endgroup#1\@@endlink}%
\providecommand \@sanitize@url [0]{\catcode `\\12\catcode `\$12\catcode
  `\&12\catcode `\#12\catcode `\^12\catcode `\_12\catcode `\%12\relax}%
\providecommand \@@startlink[1]{}%
\providecommand \@@endlink[0]{}%
\providecommand \url  [0]{\begingroup\@sanitize@url \@url }%
\providecommand \@url [1]{\endgroup\@href {#1}{\urlprefix }}%
\providecommand \urlprefix  [0]{URL }%
\providecommand \Eprint [0]{\href }%
\providecommand \doibase [0]{http://dx.doi.org/}%
\providecommand \selectlanguage [0]{\@gobble}%
\providecommand \bibinfo  [0]{\@secondoftwo}%
\providecommand \bibfield  [0]{\@secondoftwo}%
\providecommand \translation [1]{[#1]}%
\providecommand \BibitemOpen [0]{}%
\providecommand \bibitemStop [0]{}%
\providecommand \bibitemNoStop [0]{.\EOS\space}%
\providecommand \EOS [0]{\spacefactor3000\relax}%
\providecommand \BibitemShut  [1]{\csname bibitem#1\endcsname}%
\let\auto@bib@innerbib\@empty
%</preamble>
\bibitem [{\citenamefont {Abdullaev}(2006)}]{abudullaev2006}%
  \BibitemOpen
  \bibfield  {author} {\bibinfo {author} {\bibnamefont {Abdullaev},
  \bibfnamefont {S.~S.}},\ }\href@noop {} {\emph {\bibinfo {title}
  {Construction of Mappings for Hamiltonian Systems and Their Applications}}}\
  (\bibinfo  {publisher} {Spring-Vlerg},\ \bibinfo {year} {2006})\BibitemShut
  {NoStop}%
\bibitem [{\citenamefont {Biancalani}\ \emph {et~al.}(2016)\citenamefont
  {Biancalani}, \citenamefont {Bottino}, \citenamefont {Briguglio},
  \citenamefont {K?nies}, \citenamefont {Lauber}, \citenamefont {Mishchenko},
  \citenamefont {Poli}, \citenamefont {Scott},\ and\ \citenamefont
  {Zonca}}]{biancalanipop2016}%
  \BibitemOpen
  \bibfield  {author} {\bibinfo {author} {\bibnamefont {Biancalani},
  \bibfnamefont {A.}}, \bibinfo {author} {\bibnamefont {Bottino}, \bibfnamefont
  {A.}}, \bibinfo {author} {\bibnamefont {Briguglio}, \bibfnamefont {S.}},
  \bibinfo {author} {\bibnamefont {K?nies}, \bibfnamefont {A.}}, \bibinfo
  {author} {\bibnamefont {Lauber}, \bibfnamefont {P.}}, \bibinfo {author}
  {\bibnamefont {Mishchenko}, \bibfnamefont {A.}}, \bibinfo {author}
  {\bibnamefont {Poli}, \bibfnamefont {E.}}, \bibinfo {author} {\bibnamefont
  {Scott}, \bibfnamefont {B.~D.}}, \ and\ \bibinfo {author} {\bibnamefont
  {Zonca}, \bibfnamefont {F.}},\ }\href@noop {} {\bibfield  {journal} {\bibinfo
   {journal} {Phys. Plasmas.}\ }\textbf {\bibinfo {volume} {23}},\ \bibinfo
  {pages} {012108} (\bibinfo {year} {2016})}\BibitemShut {NoStop}%
\bibitem [{\citenamefont {Brizard}(1990)}]{1990brizard}%
  \BibitemOpen
  \bibfield  {author} {\bibinfo {author} {\bibnamefont {Brizard}, \bibfnamefont
  {A.~J.}},\ }\emph {\bibinfo {title} {Nonlinear Gyrokinetic Tokamak
  Physics}},\ \href@noop {} {Ph.D. thesis} (\bibinfo {year} {1990})\BibitemShut
  {NoStop}%
\bibitem [{\citenamefont {Brizard}\ and\ \citenamefont
  {Hahm}(2007)}]{2007brizard1}%
  \BibitemOpen
  \bibfield  {author} {\bibinfo {author} {\bibnamefont {Brizard}, \bibfnamefont
  {A.~J.}}\ and\ \bibinfo {author} {\bibnamefont {Hahm}, \bibfnamefont
  {T.~S.}},\ }\href@noop {} {\bibfield  {journal} {\bibinfo  {journal} {Rev.
  Mod. Phys.}\ }\textbf {\bibinfo {volume} {79}},\ \bibinfo {pages} {421}
  (\bibinfo {year} {2007})}\BibitemShut {NoStop}%
\bibitem [{\citenamefont {Cary}\ and\ \citenamefont
  {Brizard}(2009)}]{2009cary}%
  \BibitemOpen
  \bibfield  {author} {\bibinfo {author} {\bibnamefont {Cary}, \bibfnamefont
  {J.~R.}}\ and\ \bibinfo {author} {\bibnamefont {Brizard}, \bibfnamefont
  {A.~J.}},\ }\href@noop {} {\bibfield  {journal} {\bibinfo  {journal} {Rev.
  Mod. Phys.}\ }\textbf {\bibinfo {volume} {81}},\ \bibinfo {pages} {693}
  (\bibinfo {year} {2009})}\BibitemShut {NoStop}%
\bibitem [{\citenamefont {Cary}\ and\ \citenamefont
  {Littlejohn}(1983)}]{1983cary}%
  \BibitemOpen
  \bibfield  {author} {\bibinfo {author} {\bibnamefont {Cary}, \bibfnamefont
  {J.~R.}}\ and\ \bibinfo {author} {\bibnamefont {Littlejohn}, \bibfnamefont
  {R.~G.}},\ }\href@noop {} {\bibfield  {journal} {\bibinfo  {journal} {Ann.
  Phys.}\ }\textbf {\bibinfo {volume} {151}},\ \bibinfo {pages} {1} (\bibinfo
  {year} {1983})}\BibitemShut {NoStop}%
\bibitem [{\citenamefont {Chen}(2010)}]{ffchenbook}%
  \BibitemOpen
  \bibfield  {author} {\bibinfo {author} {\bibnamefont {Chen}, \bibfnamefont
  {F.~F.}},\ }\href@noop {} {\emph {\bibinfo {title} {Introduction to plasma
  physics and controlled fusion}}}\ (\bibinfo  {publisher} {Springer},\
  \bibinfo {year} {2010})\BibitemShut {NoStop}%
\bibitem [{\citenamefont {D.~Q}\ and\ \citenamefont {R}(1981)}]{1981hwang}%
  \BibitemOpen
  \bibfield  {author} {\bibinfo {author} {\bibnamefont {Hwag}, \bibfnamefont
  {D.~Q}}\ and\ \bibinfo {author} {\bibnamefont {Wilson}, \bibfnamefont {J. R}},\
  }\href@noop {} {\bibfield  {journal} {\bibinfo  {journal} {Proceedings of
  IEEE}\ }\textbf {\bibinfo {volume} {69}},\ \bibinfo {pages} {1030} (\bibinfo
  {year} {1981})}\BibitemShut {NoStop}%
\bibitem [{\citenamefont {Erckmann.V}\ and\ \citenamefont
  {Gasparino.U}(1994)}]{1994erckmannppcf}%
  \BibitemOpen
  \bibfield  {author} {\bibinfo {author} {\bibnamefont {Erckmann.V},}\ and\
  \bibinfo {author} {\bibnamefont {Gasparino.U},},\ }\href@noop {} {\bibfield
  {journal} {\bibinfo  {journal} {Plasma Phys. Contr. F.}\ }\textbf {\bibinfo
  {volume} {36}},\ \bibinfo {pages} {1869} (\bibinfo {year}
  {1994})}\BibitemShut {NoStop}%
\bibitem [{\citenamefont {Fisch}(1987)}]{1987fisch}%
  \BibitemOpen
  \bibfield  {author} {\bibinfo {author} {\bibnamefont {Fisch}, \bibfnamefont
  {N.~J.}},\ }\href@noop {} {\bibfield  {journal} {\bibinfo  {journal} {Rev.
  Mod. Phys.}\ }\textbf {\bibinfo {volume} {59}},\ \bibinfo {pages} {175}
  (\bibinfo {year} {1987})}\BibitemShut {NoStop}%
\bibitem [{\citenamefont {Hahm}(1988)}]{1988hahm}%
  \BibitemOpen
  \bibfield  {author} {\bibinfo {author} {\bibnamefont {Hahm}, \bibfnamefont
  {T.~S.}},\ }\href@noop {} {\bibfield  {journal} {\bibinfo  {journal} {Phys.
  Fluids.}\ }\textbf {\bibinfo {volume} {31}},\ \bibinfo {pages} {2670}
  (\bibinfo {year} {1988})}\BibitemShut {NoStop}%
\bibitem [{\citenamefont {Hahm}, \citenamefont {Lee},\ and\ \citenamefont
  {Brizard}(1988)}]{1988hahm2}%
  \BibitemOpen
  \bibfield  {author} {\bibinfo {author} {\bibnamefont {Hahm}, \bibfnamefont
  {T.~S.}}, \bibinfo {author} {\bibnamefont {Lee}, \bibfnamefont {W.~W.}}, \
  and\ \bibinfo {author} {\bibnamefont {Brizard}, \bibfnamefont {A.}},\
  }\href@noop {} {\bibfield  {journal} {\bibinfo  {journal} {Phys. Fluids.}\
  }\textbf {\bibinfo {volume} {31}},\ \bibinfo {pages} {1940} (\bibinfo {year}
  {1988})}\BibitemShut {NoStop}%
\bibitem [{\citenamefont {Idomura.Y}\ \emph {et~al.}(2009)\citenamefont
  {Idomura.Y}, \citenamefont {Urano.H}, \citenamefont {Aiba.N},\ and\
  \citenamefont {Tokuda.S}}]{Idomuranf2009}%
  \BibitemOpen
  \bibfield  {author} {\bibinfo {author} {\bibnamefont {Idomura.Y},}, \bibinfo
  {author} {\bibnamefont {Urano.H},}, \bibinfo {author} {\bibnamefont
  {Aiba.N},}, \ and\ \bibinfo {author} {\bibnamefont {Tokuda.S},},\ }\href@noop
  {} {\bibfield  {journal} {\bibinfo  {journal} {Nucl. Fusion}\ }\textbf
  {\bibinfo {volume} {49}},\ \bibinfo {pages} {065029} (\bibinfo {year}
  {2009})}\BibitemShut {NoStop}%
\bibitem [{\citenamefont {Lee}(1987)}]{wwleejcp1987}%
  \BibitemOpen
  \bibfield  {author} {\bibinfo {author} {\bibnamefont {Lee}, \bibfnamefont
  {W.~W.}},\ }\href@noop {} {\bibfield  {journal} {\bibinfo  {journal} {J.
  Comput. Phys.}\ }\textbf {\bibinfo {volume} {72}},\ \bibinfo {pages} {243}
  (\bibinfo {year} {1987})}\BibitemShut {NoStop}%
\bibitem [{\citenamefont {Lichtenberg.A.J}\ and\ \citenamefont
  {Lieberman.M.A}(1992)}]{lichtenberg1992}%
  \BibitemOpen
  \bibfield  {author} {\bibinfo {author} {\bibnamefont {Lichtenberg.A.J},}\
  and\ \bibinfo {author} {\bibnamefont {Lieberman.M.A},},\ }\href@noop {}
  {\emph {\bibinfo {title} {Regular and Chaotic Dynamics}}}\ (\bibinfo
  {publisher} {Springer-Verlag},\ \bibinfo {year} {1992})\BibitemShut {NoStop}%
\bibitem [{\citenamefont {Lin}, \citenamefont {Tang},\ and\ \citenamefont
  {Lee}(1995)}]{zhlinpop1995}%
  \BibitemOpen
  \bibfield  {author} {\bibinfo {author} {\bibnamefont {Lin}, \bibfnamefont
  {Z.}}, \bibinfo {author} {\bibnamefont {Tang}, \bibfnamefont {W.~M.}}, \ and\
  \bibinfo {author} {\bibnamefont {Lee}, \bibfnamefont {W.~W.}},\ }\href@noop
  {} {\bibfield  {journal} {\bibinfo  {journal} {Phys. Plasmas.}\ }\textbf
  {\bibinfo {volume} {2}},\ \bibinfo {pages} {2975} (\bibinfo {year}
  {1995})}\BibitemShut {NoStop}%
\bibitem [{\citenamefont {Littlejohn}(1983)}]{1983littlejohn}%
  \BibitemOpen
  \bibfield  {author} {\bibinfo {author} {\bibnamefont {Littlejohn},
  \bibfnamefont {R.~G.}},\ }\href@noop {} {\bibfield  {journal} {\bibinfo
  {journal} {J. Plasma Phys.}\ }\textbf {\bibinfo {volume} {29}},\ \bibinfo
  {pages} {111} (\bibinfo {year} {1983})}\BibitemShut {NoStop}%
\bibitem [{\citenamefont {Park}\ and\ \citenamefont
  {Chang}(2007)}]{gunyoungpop2007}%
  \BibitemOpen
  \bibfield  {author} {\bibinfo {author} {\bibnamefont {Park}, \bibfnamefont
  {G.}}\ and\ \bibinfo {author} {\bibnamefont {Chang}, \bibfnamefont {C.~S.}},\
  }\href@noop {} {\bibfield  {journal} {\bibinfo  {journal} {Phys. Plasmas.}\
  }\textbf {\bibinfo {volume} {14}},\ \bibinfo {pages} {052503} (\bibinfo
  {year} {2007})}\BibitemShut {NoStop}%
\bibitem [{\citenamefont {Perkins}(1977)}]{1977nfperkins}%
  \BibitemOpen
  \bibfield  {author} {\bibinfo {author} {\bibnamefont {Perkins}, \bibfnamefont
  {F.~W.}},\ }\href@noop {} {\bibfield  {journal} {\bibinfo  {journal} {Nucl.
  Fusion}\ }\textbf {\bibinfo {volume} {17}},\ \bibinfo {pages} {1197}
  (\bibinfo {year} {1977})}\BibitemShut {NoStop}%
\bibitem [{\citenamefont {Phillips}\ \emph {et~al.}(1995)\citenamefont
  {Phillips}, \citenamefont {Bell}, \citenamefont {Bell}, \citenamefont
  {Bretz}, \citenamefont {Budny}, \citenamefont {Darrow}, \citenamefont {Grek},
  \citenamefont {Hammett}, \citenamefont {Hosea}, \citenamefont {Hsuan},
  \citenamefont {Ignat}, \citenamefont {Majeski}, \citenamefont {Mazzucato},
  \citenamefont {Nazikian}, \citenamefont {Park}, \citenamefont {Rogers},
  \citenamefont {Schilling}, \citenamefont {Stevens}, \citenamefont
  {Synakowski}, \citenamefont {Taylor}, \citenamefont {Wilson}, \citenamefont
  {Zarnstorff}, \citenamefont {Zweben}, \citenamefont {Bush}, \citenamefont
  {Goldfinger}, \citenamefont {Jaeger}, \citenamefont {Murakami}, \citenamefont
  {Rasmussen}, \citenamefont {Bettenhausen}, \citenamefont {Lam}, \citenamefont
  {Scharer}, \citenamefont {Sund},\ and\ \citenamefont {Sauter}}]{1995philips}%
  \BibitemOpen
  \bibfield  {author} {\bibinfo {author} {\bibnamefont {Phillips},
  \bibfnamefont {C.~K.}}, \bibinfo {author} {\bibnamefont {Bell}, \bibfnamefont
  {M.~G.}}, \bibinfo {author} {\bibnamefont {Bell}, \bibfnamefont {R.}},
  \bibinfo {author} {\bibnamefont {Bretz}, \bibfnamefont {N.}}, \bibinfo
  {author} {\bibnamefont {Budny}, \bibfnamefont {R.~V.}}, \bibinfo {author}
  {\bibnamefont {Darrow}, \bibfnamefont {D.~S.}}, \bibinfo {author}
  {\bibnamefont {Grek}, \bibfnamefont {B.}}, \bibinfo {author} {\bibnamefont
  {Hammett}, \bibfnamefont {G.}}, \bibinfo {author} {\bibnamefont {Hosea},
  \bibfnamefont {J.~C.}}, \bibinfo {author} {\bibnamefont {Hsuan},
  \bibfnamefont {H.}}, \bibinfo {author} {\bibnamefont {Ignat}, \bibfnamefont
  {D.}}, \bibinfo {author} {\bibnamefont {Majeski}, \bibfnamefont {R.}},
  \bibinfo {author} {\bibnamefont {Mazzucato}, \bibfnamefont {E.}}, \bibinfo
  {author} {\bibnamefont {Nazikian}, \bibfnamefont {R.}}, \bibinfo {author}
  {\bibnamefont {Park}, \bibfnamefont {H.}}, \bibinfo {author} {\bibnamefont
  {Rogers}, \bibfnamefont {J.~H.}}, \bibinfo {author} {\bibnamefont
  {Schilling}, \bibfnamefont {G.}}, \bibinfo {author} {\bibnamefont {Stevens},
  \bibfnamefont {J.~E.}}, \bibinfo {author} {\bibnamefont {Synakowski},
  \bibfnamefont {E.}}, \bibinfo {author} {\bibnamefont {Taylor}, \bibfnamefont
  {G.}}, \bibinfo {author} {\bibnamefont {Wilson}, \bibfnamefont {J.~R.}},
  \bibinfo {author} {\bibnamefont {Zarnstorff}, \bibfnamefont {M.~C.}},
  \bibinfo {author} {\bibnamefont {Zweben}, \bibfnamefont {S.~J.}}, \bibinfo
  {author} {\bibnamefont {Bush}, \bibfnamefont {C.~E.}}, \bibinfo {author}
  {\bibnamefont {Goldfinger}, \bibfnamefont {R.}}, \bibinfo {author}
  {\bibnamefont {Jaeger}, \bibfnamefont {E.~F.}}, \bibinfo {author}
  {\bibnamefont {Murakami}, \bibfnamefont {M.}}, \bibinfo {author}
  {\bibnamefont {Rasmussen}, \bibfnamefont {D.}}, \bibinfo {author}
  {\bibnamefont {Bettenhausen}, \bibfnamefont {M.}}, \bibinfo {author}
  {\bibnamefont {Lam}, \bibfnamefont {N.~T.}}, \bibinfo {author} {\bibnamefont
  {Scharer}, \bibfnamefont {J.}}, \bibinfo {author} {\bibnamefont {Sund},
  \bibfnamefont {R.}}, \ and\ \bibinfo {author} {\bibnamefont {Sauter},
  \bibfnamefont {O.}},\ }\href@noop {} {\bibfield  {journal} {\bibinfo
  {journal} {Phys. Plasmas.}\ }\textbf {\bibinfo {volume} {2}},\ \bibinfo
  {pages} {2427} (\bibinfo {year} {1995})}\BibitemShut {NoStop}%
\bibitem [{\citenamefont {Qin}\ \emph {et~al.}(1999)\citenamefont {Qin},
  \citenamefont {Tang}, \citenamefont {Lee},\ and\ \citenamefont
  {Rewoldt}}]{1999honqingpop}%
  \BibitemOpen
  \bibfield  {author} {\bibinfo {author} {\bibnamefont {Qin}, \bibfnamefont
  {H.}}, \bibinfo {author} {\bibnamefont {Tang}, \bibfnamefont {W.~M.}},
  \bibinfo {author} {\bibnamefont {Lee}, \bibfnamefont {W.~W.}}, \ and\
  \bibinfo {author} {\bibnamefont {Rewoldt}, \bibfnamefont {G.}},\ }\href@noop
  {} {\bibfield  {journal} {\bibinfo  {journal} {Phys. Plasmas.}\ }\textbf
  {\bibinfo {volume} {6}},\ \bibinfo {pages} {1575} (\bibinfo {year}
  {1999})}\BibitemShut {NoStop}%
\bibitem [{\citenamefont {Stix}(1975)}]{1975stix}%
  \BibitemOpen
  \bibfield  {author} {\bibinfo {author} {\bibnamefont {Stix}, \bibfnamefont
  {T.~H.}},\ }\href@noop {} {\bibfield  {journal} {\bibinfo  {journal} {Nucl.
  Fusion}\ }\textbf {\bibinfo {volume} {15}},\ \bibinfo {pages} {737} (\bibinfo
  {year} {1975})}\BibitemShut {NoStop}%
\bibitem [{\citenamefont {Stix}(1992)}]{stixbook}%
  \BibitemOpen
  \bibfield  {author} {\bibinfo {author} {\bibnamefont {Stix}, \bibfnamefont
  {T.~H.}},\ }\href@noop {} {\emph {\bibinfo {title} {Waves in plasmas}}}\
  (\bibinfo  {publisher} {Springer-Verlag},\ \bibinfo {year}
  {1992})\BibitemShut {NoStop}%
\bibitem [{\citenamefont {Swanson}(1989)}]{swansonbook}%
  \BibitemOpen
  \bibfield  {author} {\bibinfo {author} {\bibnamefont {Swanson}, \bibfnamefont
  {D.~G.}},\ }\href@noop {} {\emph {\bibinfo {title} {Plasma waves}}}\
  (\bibinfo  {publisher} {Academic Press},\ \bibinfo {year} {1989})\BibitemShut
  {NoStop}%
\bibitem [{\citenamefont {Tronko}, \citenamefont {Bottino},\ and\ \citenamefont
  {Sonnendr¨¹cker}(2016)}]{2016tronko}%
  \BibitemOpen
  \bibfield  {author} {\bibinfo {author} {\bibnamefont {Tronko}, \bibfnamefont
  {N.}}, \bibinfo {author} {\bibnamefont {Bottino}, \bibfnamefont {A.}}, \ and\
  \bibinfo {author} {\bibnamefont {Sonnendr¨¹cker}, \bibfnamefont {E.}},\
  }\href@noop {} {\bibfield  {journal} {\bibinfo  {journal} {Phys. Plasmas.}\
  }\textbf {\bibinfo {volume} {23}},\ \bibinfo {pages} {082505} (\bibinfo
  {year} {2016})}\BibitemShut {NoStop}%
\bibitem [{\citenamefont {Tronko}\ and\ \citenamefont
  {Brizard}(2015)}]{2015tronko}%
  \BibitemOpen
  \bibfield  {author} {\bibinfo {author} {\bibnamefont {Tronko}, \bibfnamefont
  {N.}}\ and\ \bibinfo {author} {\bibnamefont {Brizard}, \bibfnamefont
  {A.~J.}},\ }\href@noop {} {\bibfield  {journal} {\bibinfo  {journal} {Phys.
  Plasmas.}\ }\textbf {\bibinfo {volume} {22}},\ \bibinfo {pages} {112507}
  (\bibinfo {year} {2015})}\BibitemShut {NoStop}%
\bibitem [{\citenamefont {Wang}(2006)}]{2006shaojiewang}%
  \BibitemOpen
  \bibfield  {author} {\bibinfo {author} {\bibnamefont {Wang}, \bibfnamefont
  {S.}},\ }\href@noop {} {\bibfield  {journal} {\bibinfo  {journal} {Phys.
  Plasmas.}\ }\textbf {\bibinfo {volume} {13}},\ \bibinfo {pages} {052506}
  (\bibinfo {year} {2006})}\BibitemShut {NoStop}%
\bibitem [{\citenamefont {Wesson}(2004)}]{wesson2004}%
  \BibitemOpen
  \bibfield  {author} {\bibinfo {author} {\bibnamefont {Wesson}, \bibfnamefont
  {J.}},\ }\href@noop {} {\emph {\bibinfo {title} {Tokamaks}}}\ (\bibinfo
  {publisher} {Oxford Press},\ \bibinfo {year} {2004})\BibitemShut {NoStop}%
\bibitem [{\citenamefont {Xu}\ and\ \citenamefont
  {Rosenbluth}(1991)}]{xuxueq1991}%
  \BibitemOpen
  \bibfield  {author} {\bibinfo {author} {\bibnamefont {Xu}, \bibfnamefont
  {X.~Q.}}\ and\ \bibinfo {author} {\bibnamefont {Rosenbluth}, \bibfnamefont
  {M.~N.}},\ }\href@noop {} {\bibfield  {journal} {\bibinfo  {journal} {Physics
  of Fluids B}\ }\textbf {\bibinfo {volume} {3}},\ \bibinfo {pages} {627}
  (\bibinfo {year} {1991})}\BibitemShut {NoStop}%
\end{thebibliography}

%

\end{document}